 \documentstyle[aas2pp4,epsf]{article}

\newcommand{\Ell}{E_\parallel}      
\newcommand{\rhoGJ}{\rho_{{\rm GJ}}}  
\newcommand{\Bp}{B_{\rm p}}        
\newcommand{\sgT}{\sigma_{\rm T}}  
\newcommand{\sgP}{\sigma_{\rm p}}  
\newcommand{\rlc}{\varpi_{\rm LC}} 
\newcommand{\Es}{\epsilon_{\rm s}} 
\newcommand{\Eg}{\epsilon_\gamma}  
\newcommand{\Emax}{\epsilon_{\rm max}}   
\newcommand{\Emin}{\epsilon_{\rm min}}   
\newcommand{\inc}{\alpha_{\rm i}}  
\newcommand{\Pcv}{P_{\rm CV}}      
\newcommand{\etaICe}{\eta_{\rm IC}^{\rm e}}
\newcommand{\etaICg}{\eta_{\rm IC}^\gamma}
\newcommand{\etaCV}{\eta_{\rm CV}}
\newcommand{\etaP}{\eta_{\rm p}}
\newcommand{\omgp}{\omega_{\rm p}} %
\newcommand{\bCM}{\beta_{\rm CM}} %

\lefthead{Hirotani, Harding, and Shibata}
\righthead{}
\begin{document}

\title{Electrodynamics of Outer-Gap Accelerator:
       Formation of Soft Power-law Spectrum
       Between 100~MeV and 3~GeV}
\author{Kouichi Hirotani, Alice K. Harding}
 \affil{Code 661.0, 
        Laboratory for High Energy Astrophysics,\\
        NASA/Goddard Space Flight Center, 
        Greenbelt, MD~20771;            \\ 
	hirotani@milkyway.gsfc.nasa.gov,
        Alice.K.Harding@nasa.gov}
\and
\author{Shinpei Shibata}
 \affil{Department of Physics, Yamagata University,
        Yamagata 990-8560, Japan;        \\ 
	shibata@sci.kj.yamagata-u.ac.jp}

\begin{abstract}
We investigate a stationary pair production cascade 
in the outer magnetosphere of a spinning neutron star.
The charge depletion due to global flows of charged particles,
causes a large electric field along the magnetic field lines.
Migratory electrons and/or positrons are accelerated by this field
to radiate gamma-rays via curvature and inverse-Compton 
processes.
Some of such gamma-rays collide with the X-rays 
to materialize as pairs in the gap.
The replenished charges partially screen the electric field, 
which is self-consistently solved 
together with the energy distribution of particles and gamma-rays
at each point along the field lines.
By solving the set of Maxwell and Boltzmann equations,
we demonstrate that an external injection of charged particles
at nearly Goldreich-Julian rate does not quench the gap 
but shifts its position
and that the particle energy distribution cannot be described
by a power-law.
The injected particles are accelerated in the gap
and escape from it with large Lorentz factors.
We show that such escaping particles migrating outside of the gap
contribute significantly to the gamma-ray luminosity for
young pulsars
and that the soft gamma-ray spectrum between 100~MeV and 3~GeV
observed for the Vela pulsar can be explained by this component.
We also discuss that the luminosity of the gamma-rays
emitted by the escaping particles 
is naturally proportional to 
the square root of the spin-down luminosity.
\end{abstract}



\section{Introduction}
\label{sec:intro}
Recent years have seen a renewal of interest in the
theory of particle acceleration in pulsar magnetospheres,
after the launch of the {\it Compton Gamma-ray Observatory}
(e.g., for the Vela pulsar, Kanbach et al. 1994, Fierro et al. 1998;
 for PSR B1706--44, Thompson et al. 1996;
 for Geminga, Mayer-Hasselwander et al. 1994, Fierro et al. 1998; 
 for PSR B1055--52, Thompson et al. 1999).
The modulation of the $\gamma$-ray light curves 
testifies to the particle acceleration either at the polar cap 
(Harding, Tademaru, \& Esposito 1978; Daugherty \& Harding 1982, 1996;
 Sturner, Dermer, \& Michel 1995),
or at the vacuum gaps in the outer magnetosphere
(Cheng, Ho, \& Ruderman 1986a,b, hereafter CHR;
 Chiang \& Romani 1992, 1994; Romani \& Yadigaroglu 1995;
 Higgins \& Henriksen 1997, 1998).
Both models predict that electrons and positrons are
accelerated in a charge depletion region, a potential gap,
by the electric field along the magnetic field lines
to radiate high-energy $\gamma$-rays via the curvature process.
However, there is an important difference between these two models:
An polar-gap accelerator releases very little angular momenta,
while an outer-gap one could radiate them efficiently.
In addition, three-dimensional outer-gap models
commonly explain double-peak light curves with strong bridges
observed for $\gamma$-ray pulsars.
On these grounds, the purpose of the present paper 
is to explore further into the analysis of the outer-gap accelerator.

In the CHR picture, 
the gap is assumed to be geometrically thin in the transfield
direction on the poloidal plane
in the sense $D_\perp \ll W$,
where $D_\perp$ represents the typical transfield thickness of the gap,
while $W$ does the width along the magnetic field lines.
In this limit, the acceleration electric field is partially screened 
by the zero-potential walls separated with a small distance $D_\perp$;
as a result, the gap, which is assumed to be vacuum,
extends from the null surface to (the vicinity of) 
the light cylinder. 
Here, the null surface is defined as the place on which
the local Goldreich-Julian charge density
\begin{equation}
  \rhoGJ \equiv -\frac{\Omega B_z(s,z)}{2\pi c}
  \label{eq:def_rhoGJ}
\end{equation}
vanishes, 
where $\Omega$ refers to the angular frequency of the neutron star,
$B_z$ the magnetic field component
projected along the rotational axis,
and $c$ the velocity of light;
$s$ and $z$ refer to the coordinates parallel and 
perpendicular, respectively, to the poloidal magnetic field. 
The star surface corresponds to $s=0$;
$s$ increases outwardly along the field lines.
The last-open field line corresponds to $z=0$; 
$z$ increases towards the magnetic axis (in the same hemisphere).

If $B_z>0$ holds in the starward side of the null surface,
a positive acceleration field arises in the gap.
The light cylinder is defined as the surface
where the azimuthal velocity of a plasma would coincide with $c$
if it corotated with the magnetosphere.
Its radius from the rotational axis becomes the 
so-called \lq light cylinder radius',
$\rlc \equiv c / \Omega$.
Particles are not allowed to migrate inwards beyond this surface
because of the causality in special relativity.

It should be noted that the null surface is not a special place
for the gap electrodynamics in the sense that the plasmas are not
completely charge-separated in general and that the particles
freely pass through this surface inwards and outwards.
Therefore, the gap inner boundary is located near to the null surface,
not because particle injection is impossible across this surface
(as previously discussed),
but because the gap is vacuum and transversely thin.

Then what occurs in the CHR picture if the gap becomes 
no longer vacuum?
To consider this problem rigorously, we have to examine
the Poisson equation for the electrostatic potential. 
In fact, as will be explicitly demonstrated in the next section,
the original vacuum solution obtained in the pioneering work by
CHR cannot be applied to a non-vacuum CHR picture.
We are, therefore,  motivated by the need to solve
self-consistently the Poisson equation together with the
Boltzmann equations for particles and $\gamma$-rays.
Although the ultimate goal is to solve three-dimensional issues,
a good place to start is to examine one-dimensional problems.
In this context,
Hirotani and Shibata (1999a,~b,~c; hereafter Papers~I,~II,~III) 
first solved the Boltzmann equations
together with the Maxwell equations
one-dimensionally along the field lines,
extending the idea originally developed for 
black-hole magnetospheres by Beskin et al. (1992).
In Paper~I, II, and III, 
they assumed that the gap is 
geometrically thick in the transfield direction
in the sense $D_\perp \gg W$.

There is one important finding in this second picture:
The gap position shifts if there is a particle injection 
across either of the boundaries 
(Hirotani \& Shibata 2001, 2002a,b; hereafter Papers~VII, VIII, IX).
For example, 
when the injection rate across the outer (or inner) boundary
becomes comparable to the typical Goldreich-Julian value, 
the gap is located close to the neutron star surface 
(or to the light cylinder).
In other words, an outer gap is not quenched even 
when the injection rate of a completely charge-separated
plasma across the boundaries approaches the 
typical Goldreich-Julian value. 
Thus, an outer gap can coexist with a polar-cap accelerator;
this forms a striking contrast to the first, CHR picture.
It is also found in the second picture 
that an outer gap is quenched if the 
{\it created} particle density within the gap
exceeds several percent of the Goldreich-Julian value.
That is, the {\it discharge} of created pairs is essential to screen
the acceleration field.

The purpose of this paper is to examine the second picture
more closely.
In all the previous works in the second picture,
the particle energy distribution has been assumed to be
mono-energetic in the sense that the particles attain
the equilibrium Lorentz factor at each point,
in a balance between the electrostatic acceleration
and the radiation-reaction forces.
In the present paper, we discard this assumption
and explicitly consider the energy dependence of
particles by solving the Boltzmann equations for
positrons and electrons. 
We will demonstrate that the particle energy
distribution cannot be represented 
either by a power law or by the mono-energetic approximation.
We will further show that a soft power-law spectrum
is generally formed in 100~MeV-GeV energies
as a result of the superposition of the 
curvature spectra emitted by particles 
migrating at different positions.

In the next section, 
we describe the difficulties of electrodynamics
found in the first picture.
We then present the basic equations in \S~\ref{sec:basic}.
and apply the theory to four
$\gamma$-ray pulsars and compare the predictions with observations
in \S~\ref{sec:app}.
In the final section, 
we discuss the possibilities of the unification of our picture
with the CHR picture,
as well as the unification of the outer-gap and polar-cap models.

\section{Difficulties in Previous Outer-gap Models}
\label{sec:diffic}
To elucidate the electrodynamic difficulties in previous outer-gap
models, we have to examine the Poisson equation for the 
electrostatic potential, $\Psi$.
Since the azimuthal dimension is supposed to be large
compared with $D_\perp$ in conventional outer-gap models, 
the Poisson equation is reduced to the following 
two-dimensional form on the poloidal plane
\begin{equation} 
  -\frac{\partial^2 \Psi}{\partial s^2}
  -\frac{\partial^2 \Psi}{\partial z^2}
  = 4 \pi [\rho_{\rm e}(s,z)-\rhoGJ(s,z)].
  \label{eq:Poisson_2D}
\end{equation}
The true charge density, $\rho_{\rm e}$, 
is given by $\rho_{\rm e}= \rho_+ +\rho_-$, 
where $\rho_+$ and $\rho_-$ represent
the positronic and electronic charge densities, respectively.
In the CHR picture, the magnetic field is supposed to have a
single-signed curvature.
As a result,
the particle number density
$(\rho_+ -\rho_-)/e$ grows exponentially in $z$ direction.
Because of this exponential growth of the particle number density,
it has been considered that most of the $\gamma$-rays are emitted
from the higher altitudes (i.e., large $z$ regions) in the gap.

To explain the observed $\gamma$-ray luminosities with a small $D_\perp$,
one should assume that the created current density 
becomes the typical Goldreich-Julian value in the higher altitudes. 
That is, the conserved current density 
per magnetic flux tube with strength $B$, should satisfy
\begin{equation}
  \frac{c\rho_+}{B} +\frac{-c\rho_-}{B}
  \sim \frac{\Omega}{2\pi}
  \label{eq:consv_0}
\end{equation}
in the order of magnitude.
However, such a copious pair production will screen the 
local acceleration field, 
$-\partial\Psi/\partial s$, 
as the Poisson equation~(\ref{eq:Poisson_2D}) indicates.

This screening effect is particularly important near to the
inner boundary.
Without loss of any generality, 
we can assume $-\partial\Psi/\partial s>0$.
In this case, because of the discharge, 
only electrons exist at the inner boundary. 
(We may notice here that external particle injections are not 
considered in the CHR picture.)
Thus, we obtain
$\rho_{\rm e}/B = \rho_-/B \sim -\Omega/(2\pi c)$
in the order of magnitude.
In the vicinity of the inner boundary,
we can Fourier-analyze equation~(\ref{eq:Poisson_2D})
in $z$ direction to find out that the $-\partial^2\Psi/\partial z^2$ term
contributes only to reduce 
$\partial(-\partial\Psi/\partial s)/\partial s$,
the $s$ gradient of the acceleration field.
Thus, a positive $-\rhoGJ$ must cancel the negative 
$\rho_{\rm e}$ to make the right-hand side be positive.
That is, at the inner boundary, 
\begin{equation}
  -\frac{\rhoGJ}{B}
  = \frac{\Omega}{2\pi c} \frac{B_z}{B}
  > \frac{\vert \rho_{\rm e} \vert}{B}
  \sim \frac{\Omega}{2\pi c}
  \label{eq:inBDcond}
\end{equation}
must be satisfied,
so that the acceleration field may not change sign in the gap.
It follows that the polar cap, where $B_z \sim B$ holds,
is the only place for the inner boundary of the \lq outer' gap 
to be located,
if the created particle number density in the gap is comparable to the
typical Goldreich-Julian value (eq.~[\ref{eq:consv_0}]). 
Such a non-vacuum gap must extend from the polar cap
(not from the null surface where $\rhoGJ$ vanishes)
to the light cylinder.
We can therefore conclude that 
the original vacuum solution obtained by
CHR cannot be applied to a non-vacuum CHR picture 
when there is a sufficient pair production that is needed 
to explain the observed $\gamma$-ray luminosity.

To construct a self-consistent model,
we have to solve equation~(\ref{eq:Poisson_2D})
together with the
Boltzmann equations for particles and $\gamma$-rays. 
We formulate them in the next section.

\section{Basic Equations}
\label{sec:basic}
%
%
\subsection{Poisson equation}
\label{sec:poisson}
Neglecting relativistic effects,
and assuming that typical transfield thickness, $D_\perp$,
is greater than or comparable to the longitudinal width, $W$,
we can reduce the Poisson equation for non-corotational potential
$\Psi$ into the one-dimensional form 
(Hirotani 2000b, Paper~VI; see also \S~2 in Michel 1974)
\begin{equation}
 -\nabla^2 \Psi 
    = -\frac{d^2\Psi}{ds^2}+\frac{\Psi}{D_\perp^2}
    = 4\pi \left[ \rho_{\rm e}(s) +\frac{\Omega B_z(s)}{2\pi c} 
           \right].
  \label{eq:Poisson_1}
\end{equation}

As described at the end of \S~3 in Paper~VII, 
it is convenient to introduce the 
Debye scale length $c/\omega_{\rm p}$, 
where
\begin{equation}
  \omega_{\rm p} = \sqrt{ \frac{4\pi e^2}{m_{\rm e}}
	                  \frac{\Omega B^{\rm in}}{2\pi ce} },
  \label{eq:def-omegap}
\end{equation}
and $B^{\rm in}$ is the magnetic field strength at the inner boundary;
$e$ designates the magnitude of the charge on an electron,
$m_{\rm e}$ the rest mass of an electron.
Thus, we can introduce the following dimensionless coordinate variable:
\begin{equation}
  \xi 
  \equiv \frac{\omega_{\rm p}}{c} s
  = 1.87 \times 10^5 
      \Omega_2^{-1/2}
      \left(\frac{B^{\rm in}}{10^5\mbox{G}}\right)^{1/2}
      \frac{s}{\rlc}.
  \label{eq:def-xi}
\end{equation}
where $\Omega_2 \equiv \Omega/(10^2 \mbox{rad s}^{-1})$.

By using such dimensionless quantities, we can rewrite
the equation~(\ref{eq:Poisson_1}) into
\begin{equation}
  \Ell = -\frac{d\psi}{d\xi}
       = -\frac{d\Psi}{ds} \cdot \frac{c}{\omgp}
          \frac{e}{m_{\rm e}c^2},
  \label{eq:basic-1}
\end{equation}
\begin{equation}
  \frac{d\Ell}{d\xi}
  =-\frac{\psi}{\Delta_\perp^2}
   +\frac{B(\xi)}{B^{\rm in}} 
    \left[ \int_1^\infty n_+ d\Gamma -\int_1^\infty n_- d\Gamma
    \right]
    + \frac{B_z(\xi)}{B^{\rm in}}
  \label{eq:basic-2}
\end{equation}
where 
\begin{equation}
  \psi(\xi) \equiv \frac{e\Psi(s)}{m_{\rm e}c^2},
  \quad
  \Delta_\perp \equiv \frac{D_\perp}{c/\omega_{\rm p}},
  \label{eq:def_Dperp}
\end{equation}
and the particle distribution functions are defined by
\begin{equation}
  n_\pm(\xi,\Gamma) \equiv 
    \frac{2\pi ce}{\Omega} \frac{N_\pm(s,\Gamma)}{B(\xi)};
  \label{eq:def-n}
\end{equation}
$N_+(s,\Gamma)$ and $N_-(s,\Gamma)$ 
represent the distribution functions
of positrons and electrons, respectively,
at position $s$ and Lorentz factor $\Gamma$.
We evaluate the dimensionless Goldreich-Julian charge density
$B_z/B^{\rm in}$ in equation~(\ref{eq:basic-2}) at each $s$,
by using the Newtonian dipole field
(see eqs.[9]-[12] in Paper~VIII for details).

\subsection{Particle Boltzmann Equations}
\label{sec:boltz_part}
On the poloidal plane, particles migrate along the magnetic field lines.
Therefore, at time $t$, 
the distribution function $N$ of particles
obeys the following Boltzmann equation,
\begin{equation}
  \frac{\partial{N}}{\partial t}
    + \frac{\mbox{\boldmath$p$}}{m_{\rm e}\Gamma} 
      \cdot \mbox{\boldmath$\nabla$} N
    + \mbox{\boldmath$F$}_{\rm ext} \cdot 
      \frac{\partial N}{\partial \mbox{\boldmath$p$}}
  = S(t,s,\mbox{\boldmath$p$}),
  \label{eq:boltz_1}
\end{equation}
where 
{\boldmath$p$} refers to the particle momentum,
$\mbox{\boldmath$F$}_{\rm ext}$ 
the external forces acting on particles,
and $S$ the collision terms.
In the present paper, 
$\mbox{\boldmath$F$}_{\rm ext}$ consists of the Lorentz
and the curvature radiation reaction forces.
Since the magnetic field is much less than the critical value 
($4.41 \times 10^{13}$~G),
quantum effects can be neglected in the outer magnetosphere.
As a result, curvature radiation takes place continuously and 
can be regarded as an external force acting on a particle.
If we instead put the collision term associated with 
the curvature process in the right-hand side, 
the energy transfer in each collision would be too small
to be resolved by the energy grids
(see description below eq.~[\ref{eq:def_etaIC_0}] for details).
We take the $\gamma$-ray production rate due to
curvature process into account 
consistently in the $\gamma$-ray Boltzmann equations.

\subsubsection{Stationary Boltzmann Equations}
\label{sec:stationary}
Imposing a stationary condition 
\begin{equation}
  \frac{\partial}{\partial t} 
  + \Omega_{\rm p} \frac{\partial}{\partial \phi} = 0,
  \label{eq:stationary}
\end{equation}
and neglecting the pitch-angle dependence, 
we can reduce equation~(\ref{eq:boltz_1}) to 
(Appendix~\ref{sec:red_boltz}) 
\begin{equation}
   \frac{\partial n_+}{\partial\xi}
   +\left[ \Ell 
          -\frac{\Pcv(\xi,\Gamma)}{m_{\rm e}c^2\omgp} \right]
    \frac{\partial n_+}{\partial\Gamma}
  =  S_+(\xi,\Gamma),
 \label{eq:boltz_2}
\end{equation}
\begin{equation}
  \frac{\partial n_-}{\partial\xi}
   -\left[ \Ell 
          -\frac{\Pcv(\xi,\Gamma)}{m_{\rm e}c^2\omgp} \right]
    \frac{\partial n_-}{\partial\Gamma}
  = -S_-(\xi,\Gamma),
 \label{eq:boltz_3}
\end{equation}
where $\Omega_{\rm p}$ designates the angular velocity of
particles due to 
$\mbox{\boldmath$E$} \times \mbox{\boldmath$B$}_{\rm p}$ drift.
Since the drift motion due to the gradient and the curvature of 
$\mbox{\boldmath$B$}_{\rm p}$ 
is negligible for typical outer-gap parameters,
$\Omega_{\rm p}$ coincides $\Omega$
provided that $B_\phi=0$ and
$\vert\Ell\vert \ll \vert E_\perp \vert = B_{\rm p}\varpi/\rlc$ hold,
where $\varpi$ refers to the distance from the rotation axis.
Radiation-reaction force due to the curvature radiation,
$\Pcv/c$, is given by 
\begin{equation}
  \frac{\Pcv}{c}
  = \frac{2e^2 \Gamma^4}{3\rho_{\rm c}^2}
  \label{def_Pcv}
\end{equation}
where $\rho_{\rm c}(s)$ refers to the curvature radius of the
magnetic field line.
For a justification of adopting the pure-curvature formula,
see \S~\ref{sec:small_pitch}.

\subsubsection{Collision terms}
\label{sec:coll_terms}
We assume in this paper that $\gamma$-rays are either outwardly or
inwardly propagating along the local magnetic field lines.
For simplicity, we neglect the deviation of the $\gamma$-rays
from the field lines due to magnetic curvature.
In what follows, we assume that the soft photons are emitted from the 
neutron star and hence unidirectional at the gap.
Then the cosine of the collision angle $\mu$ has a unique value
$\mu_+$ or $\mu_-$, for outwardly or inwardly propagating
$\gamma$-rays, respectively;
$\mu_+$ and $\mu_-$ are determined by the
magnetic inclination $\inc$ at each position $\xi$.
Under these assumptions,
the source term can be expressed as
\begin{eqnarray}
  \lefteqn{ \omgp S_+(\xi,\Gamma) 
      = - \int_{\Eg<\Gamma} d\Eg \etaICg(\Eg,\Gamma,\mu_+) 
                                 n_+(\xi,\Gamma)}
  \nonumber \\  
  &+& \int_{\Gamma_i>\Gamma} d\Gamma_i \, 
           \etaICe (\Gamma_i, \Gamma, \mu_+) n_+(\xi,\Gamma_i)
  \nonumber \\  
  &+& \frac{B^{\rm in}}{B(\xi)}
      \int d\Eg \left[ \frac{\partial \etaP(\Eg,\Gamma,\mu_+)}
                            {\partial \Gamma} g_+
                      +\frac{\partial \etaP(\Eg,\Gamma,\mu_-)}
                            {\partial \Gamma} g_-
                \right]
  \nonumber \\  
   \label{eq:src_1}
\end{eqnarray}
\begin{eqnarray}
  \lefteqn{ \omgp S_-(\xi,\Gamma) 
      = - \int_{\Eg<\Gamma} d\Eg \etaICg(\Eg,\Gamma,\mu_-) 
                                 n_-(\xi,\Gamma)}
  \nonumber \\  
  &+& \int_{\Gamma_i>\Gamma} d\Gamma_i \, 
           \etaICe (\Gamma_i, \Gamma, \mu_-) n_-(\xi,\Gamma_i)
  \nonumber \\  
  &+& \frac{B^{\rm in}}{B(\xi)}
      \int d\Eg \left[ \frac{\partial \etaP(\Eg,\Gamma,\mu_+)}
                            {\partial \Gamma} g_+
                      +\frac{\partial \etaP(\Eg,\Gamma,\mu_-)}
                            {\partial \Gamma} g_-
                \right]
  \nonumber \\  
   \label{eq:src_2}
\end{eqnarray}
where the dimensionless $\gamma$-ray distribution function
$g_\pm$ are defined as
\begin{equation}
  g_\pm(\xi,\Eg) \equiv \frac{2\pi ce}{\Omega B^{\rm in}}
                        G_\pm(s,\Eg);
  \label{eq:def_g1}
\end{equation}
$G_\pm$ refers to the number of $\gamma$-rays per unit volume
per unit dimensionless energy $\Eg=h\nu_\gamma/m_{\rm e}c^2$.
We  will explicitly define the differential redistribution function
for the pair production, $\partial \etaP /\partial\Gamma$,
by equation~(\ref{eq:def_etaP_2}),
after briefly describing the soft photon field in 
equations~(\ref{eq:soft_flux})-(\ref{eq:def_EsAST}).

If we multiply $d\Gamma$ on both sides of equation~(\ref{eq:src_1})
or (\ref{eq:src_2}),
the first (or the second) term in the right-hand side
represents the rate of particles 
disappearing from (or appearing into) the energy interval 
$m_{\rm e}c^2 \Gamma$ and $m_{\rm e}c^2 (\Gamma+d\Gamma)$
due to inverse-Compton (IC) scatterings;
the third (or the fourth) terms does the rate of particles created 
via collisions between the outgoing (or the ingoing) $\gamma$-rays and 
ambient soft photons.
The positrons (or electrons) are supposed to
collide with the soft photons at the same angle as the
outwardly (or inwardly) propagating $\gamma$-rays;
therefore, the same collision angle $\cos^{-1}\mu_+$ (or $\cos^{-1}\mu_-$)
is used for both IC scatterings and pair production in 
equations (\ref{eq:src_1}) and (\ref{eq:src_2}).
The collisions tend to be head-on (or tail-on) 
for inwardly (or outwardly) propagating $\gamma$-rays, 
as the gap approaches the star.

The IC redistribution function 
$\etaICg(\Eg,\Gamma,\mu)$ represents the probability
that a particle with Lorentz factor $\Gamma$ upscatters photons 
into energies between $\Eg$ and $\Eg+d\Eg$ per unit time when the 
collision angle is $\cos^{-1}\mu$.
On the other hand, $\etaICe(\Gamma_i,\Gamma,\mu)$ describes
the probability that a particle changes Lorentz factor from
$\Gamma_i$ to $\Gamma$ in a scattering.
Thus, energy conservation gives
\begin{equation}
  \etaICe(\Gamma_i,\Gamma_f,\mu) 
     = \etaICg(\Gamma_i-\Gamma_f,\Gamma_i,\mu)
  \label{eq:rel_etaIC}
\end{equation}

We introduce the following dimensionless IC redistribution functions,
\begin{equation}
  \etaICg{}_\pm{}_{,i}(l)
  \equiv \frac1{\omgp}
         \int_{b_{i-1}}^{b_i}
            \etaICg(\Eg,b_l,\mu_\pm)d\Eg,
  \label{eq:def_etaIC_0}
\end{equation}
where the particle Lorentz factor is discretized as $\Gamma=b_l$;
the $\gamma$-ray energy is divided into $-l_0+1+l_m=15+l_m$ bins as
$b_{l_0}= b_{-14}= 10^{-15/4}\times b_1$,
$b_{-13}= 10^{-14/4}\times b_1$,
$b_{-12}= 10^{-13/4}\times b_1$,
$\ldots$,
$b_{-2}= 10^{-3/4}\times b_1$,
$b_{-1}= 10^{-2/4}\times b_1$,
$b_{ 0}= 10^{-1/4}\times b_1$,
$b_{ 1}= 10^5$,
$b_{ 2}= b_1+\Delta b$,$\ldots$,
$b_{ i}= b_1+(i-1)\Delta b$,$\ldots$,
$b_{l_m}= b_1+(l_m-1)\Delta b$.
We normally use $l_m=256$ and $\Delta b=1.5 \times 10^5$.
In general, $\etaICg{}_\pm{}_{,i}(l)$ can be defined by 
the soft photon flux $dF_{\rm s}/d\Es$ and 
the Klein-Nishina cross section $\sigma_{\rm KN}$ as follows: 
\begin{eqnarray}
  \lefteqn{\etaICg{}_\pm{}_{,i}(l) 
           = \frac{1-\beta\mu_\pm}{\omgp}}
  \nonumber \\
  &\times&
      \int_{\Emin}^{\Emax} d\Es \frac{dF_{\rm s}}{d\Es}
      \int_{b_{i-1}}^{b_i} d\Eg \frac{d{\Eg}^\ast}{d\Eg}
      \int_{-1}^{1} d\Omega_\gamma^\ast 
           \frac{d\sigma_{\rm KN}^\ast}{d{\Eg}^\ast d\Omega_\gamma^\ast}
  \nonumber \\
  \label{eq:def_etaICg_1}
\end{eqnarray}
where $\beta \equiv \sqrt{1-1/\Gamma^2}$ is virtually unity, 
$\Omega_\gamma$ the solid angle of upscattered photon,
the asterisk denotes the quantities in the electron (or positron) 
rest frame.
In the rest frame of a particle,
a scattering always takes place well above the resonance energy.
Thus, the classical formula of the Klein-Nishina cross section
can be applied for the present problem.
The soft photon flux per unit dimensionless
photon energy $\Es$ [$s^{-1}\mbox{cm}^{-2}$] 
is written as $dF_{\rm s}/d\Es$.

We assume in this paper that soft photons are supplied by the
thermal radiation from the neutron star.
If a soft emission component cannot be represented by
a black body spectrum, we use the observed spectrum
of $dF_{\rm s}/d\Es$ corrected for the interstellar absorption.
If it can be represented by a blackbody component,
we adopt
\begin{equation}
  \frac{dF_{\rm s}}{d\Es}
  = \pi \frac{B_{\rm s}(T)}{\Es}
    \left( \frac{r_{\rm NS}}{r} \right)^2,
  \label{eq:soft_flux}
\end{equation}
where $r$ and $r_{\rm NS}$ refers to the distance from the star center
and the neutron star radius, respectively;
the Planck function 
[$\mbox{ergs s}^{-1} \mbox{cm}^{-2} \mbox{ster}^{-1} \mbox{ergs}^{-1}$]
is related with the blackbody temperature 
($\delta \equiv kT/m_{\rm e}c^2$) as
\begin{equation}
  B_{\rm s}(T)
  = \frac{2(m_{\rm e}c^2)^3}{c^2h^3} 
    \frac{\Es^3}{\exp(\Es/\delta)-1}.
  \label{eq:planck}
\end{equation}
Substituting equations~(\ref{eq:soft_flux}) into (\ref{eq:def_etaICg_1}),
and executing the integration over $\Eg$, we obtain (Appendix~B)
\begin{eqnarray}
  \lefteqn{\etaICg{}_\pm{}_{,i}(l)
           = \frac{2\pi^2 e^4 m_{\rm e}}{h^3 \omgp}
             (1-\beta\mu_\pm)
             \left( \frac{r_{\rm NS}}{r} \right)^2}
  \nonumber \\
  &\times&
    \sum_j A_j
       \int_{\Emin}^{\Emax} d\Es 
           \frac{{\Es}^2}{\exp(\Es/\delta_j)-1}
       \int_{-1}^{1} dx^\ast f_{\rm IC},
  \nonumber \\
  \label{eq:def_etaIC_2}
\end{eqnarray}
where 
\begin{equation}
  f_{\rm IC} \equiv \left\{
  \begin{array}{rl}
    z^2[z+1/z+(x^\ast)^2-1], 
      & \quad \mbox{for $z_{i-1}<z<z_i$} \\
    0 & \quad \mbox{otherwise};
  \end{array} \right.
  \label{eq:def_fIC}
\end{equation}
\begin{equation}
  z \equiv \frac{1}{1+{\Es}^\ast(1-x^\ast)}
  \label{eq:def_z}
\end{equation}
\begin{equation}
  z_i \equiv \frac{b_i}{\Gamma(1+\beta x^\ast \mu^\ast)\Es{}^\ast}
           = \frac{b_i}{b_l   (1+\beta x^\ast \mu^\ast)\Es{}^\ast}
  \label{eq:def_zi}
\end{equation}
$\mu^\ast$ and ${\Es}^\ast$ are related with $\mu$ and $\Es$
by the Lorentz transformation as
\begin{equation}
  \mu^\ast = \frac{\mu-\beta}{1-\beta \mu},
  \quad
  {\Es}^\ast = \Gamma(1-\beta\mu)\Es.
  \label{eq:def_EsAST}
\end{equation}
Since $\etaICg{}_{\pm,i}(l)$ is defined at $\Gamma=b_l$,
$\Gamma$ is replaced with $b_l$ in equation~(\ref{eq:def_zi}).
Moreover, $A_j$ represents the observed emission area of the 
$i$th blackbody component normalized by $4 \pi r_{\rm NS}^2$. 
For example, if a soft photon field consists of 
the whole neutron star surface emission with temperature $T_{\rm s}$
and a heated polar cap emission with area $\pi r_h^2$ and 
temperature $T_{\rm h}$, we obtain 
$A_1=1.0$, $\delta_1= kT_{\rm s}/m_{\rm e}c^2$ and
$A_2=\pi r_h^2/(4\pi r_{\rm NS}^2)$, $\delta_2= kT_{\rm h}/m_{\rm e}c^2$.

To define a dimensionless pair-production redistribution function,
we consider the case when a $\gamma$-ray photon with energy 
$m_{\rm e}c^2 b_i$ materializes as a pair with Lorentz factor $b_l$.
Dividing the pair production rate [$s{}^{-1}$] per Lorentz factor
by $\omgp$, we obtain
\begin{eqnarray}
  \etaP{}_\pm{}_{,l}(i)
  &\equiv& \frac{1}{\omgp} 
           \frac{\partial\etaP}{\partial\Gamma}(b_i,b_l,\mu_\pm)
  \nonumber \\
  &=& \frac{1-\mu_\pm}{\omgp}
      \int_{\epsilon_{\rm th}}^\infty d\Es
         \frac{dF_{\rm s}}{d\Es}
         \frac{d\sgP}{d\Gamma},
  \label{eq:def_etaP_2}
\end{eqnarray}
where the pair-production threshold energy is defined by
\begin{equation}
  \epsilon_{\rm th} \equiv \frac{2}{1-\mu}\frac{1}{\Eg}.
  \label{eq:def_Eth}
\end{equation}
The collision angles $\cos^{-1}\mu_\pm$ are determined 
by the geometry, in the same manner as $\etaICg{}_{\pm,i}$.
The differential cross section for pair production is given by
\begin{eqnarray}
  \lefteqn{\frac{d\sgP}{d\Gamma}
    = \frac38 \sgT \frac{1-\bCM^2}{\Eg}}
  \nonumber\\
  &\times&
    \left[ \frac{1+\bCM^2(2-\mu_{\rm CM}^2)}{1-\bCM^2\mu_{\rm CM}^2}
          -\frac{2\bCM^4(1-\mu_{\rm CM}^2)^2}{(1-\bCM^2\mu_{\rm CM}^2)^2}
    \right],
  \label{eq:def_sgP}
\end{eqnarray}
where $\sgT$ refers to the Thomson cross section and
the center-of-mass quantities are defined as
\begin{equation}
  \mu_{\rm CM} \equiv \pm \frac{2\Gamma-\Eg}{\bCM \Eg}, \quad
  \bCM^2 \equiv 1-\frac{1}{(1-\mu)\Es\Eg},
\end{equation}
where $\mu=\mu_+$ (or $\mu_-$) for $\etaP{}_{+,l}$ 
(or $\etaP{}_{-,l}$). 

Let us now discretize the differential equations.
Denoting $f_l^k$ as the quantity $f$ evaluated at
$\xi=\xi^k$ and $\Gamma= b_l$ (or $\Eg= b_l$),
the source terms (eq.~[\ref{eq:src_1}] and [\ref{eq:src_2}]) can be
represented as
\begin{eqnarray}
  {S_\pm}_l^k 
  &=& - \sum_{i<l} {\etaICg}_{\pm,i}(  l) \cdot n_\pm{}_l^k
            + \sum_{i>l} {\etaICg}_{\pm,i}(i-l) n_\pm{}_i^k
  \nonumber \\  
        &+& \frac{B^{\rm in}}{B^k}
            \sum_{i} \left[ \etaP{}_{+,l}(i) g_+{}_{,i}^k
                           +\etaP{}_{-,l}(i) g_-{}_{,i}^k
                     \right].
   \label{eq:src_3}
\end{eqnarray}
The dimensionless $\gamma$-ray distribution functions are defined
above the energy $m_{\rm e}c^2 b_{l_0}= 5.11 \times 10^7$~eV as
\begin{equation}
  g_\pm{}_{,i}^k 
  \equiv \frac{2\pi ce}{\Omega B^{\rm in}}
         \int_{b_{i-1}}^{b_i} 
            d\Eg G_\pm\left(\frac{c}{\omgp}\xi^k,\Eg \right),
  \label{eq:def_g2}
\end{equation}
where $i=l_0+1,\ldots,-1,0,1,2,\ldots,l_m$.
On the other hand, the particle distribution functions are defined
above $\Gamma \geq b_1 = 10^5$ as
\begin{equation}
  n_\pm{}_{,i}^k 
  \equiv \frac{2\pi ce}{\Omega B^k}
         N_\pm\left(\frac{c}{\omgp}\xi^k,b_l \right),
  \label{eq:def_n}
\end{equation}
where $l=1,2,\ldots,l_m$.
Note that $n_\pm$ are normalized by $B$ at each point $\xi^k$,
while $g_\pm$ are by $B(\xi^{\rm in})$.

Since the IC scatterings do not alter the particle number,
the sum of the first and the second terms in the right-hand 
side of equation~(\ref{eq:src_3}) over $l$, vanishes identically.
Therefore, 
\begin{equation}
  \sum_l \left( S_+{}_l^k -S_-{}_l^k \right) = 0
  \label{eq:sum_src}
\end{equation}
holds, because the created number of positrons is identical to 
that of electrons.

\subsubsection{Current Conservation}
\label{sec:conserv}
\begin{figure} 
\centerline{ \epsfxsize=9cm \epsfbox[-70 0 400 200]{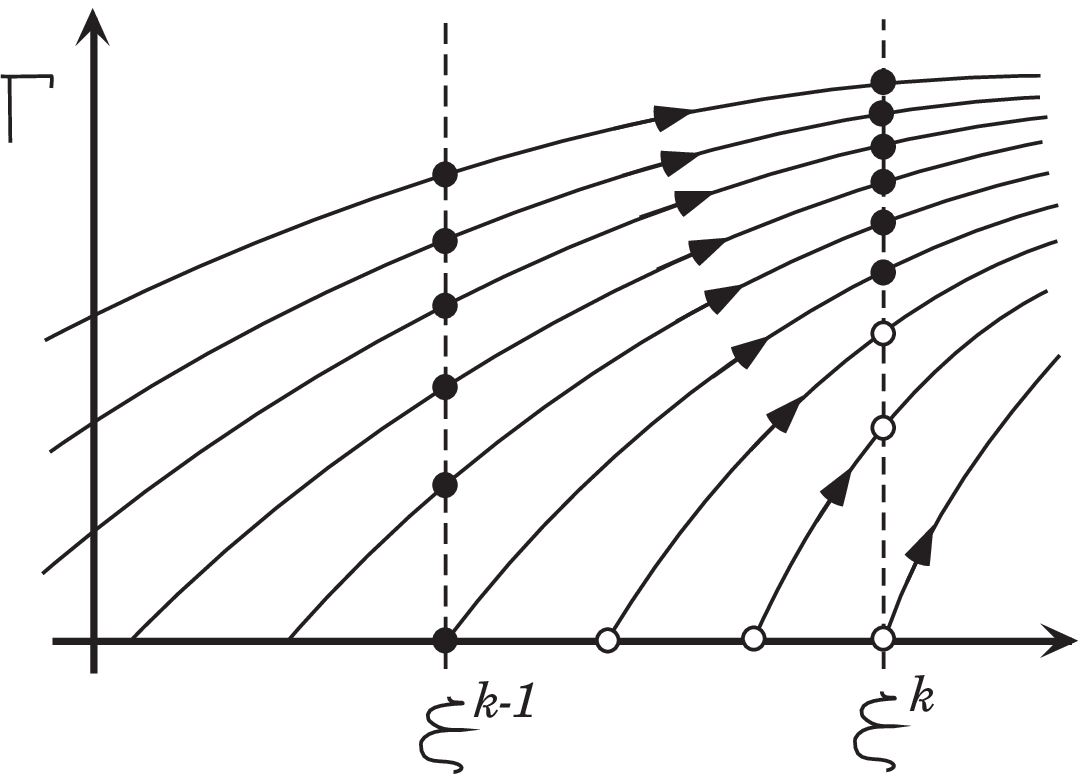} } 
\caption{
Schematic picture of characteristics in the phase space ($\xi$,$\Gamma$),
where the dimensionless coordinate $\xi$ is related with $s$ 
by equation~(\ref{eq:def-xi}).
Positrons (or electrons) move along these characteristics 
when $\Ell>0$ (or $\Ell<0$).
The filled circles indicate the particles moved from $\xi=\xi^{k-1}$
to $\xi^k$ in a certain time interval.
Created particles (open circles) are assumed to migrate immediately
outwards after the birth.
        }
\label{fig:char}
\end{figure}

Let us, for the moment, consider the current conservation problem.
Since we are considering a pure curvature process
(i.e., neglecting the pitch-angle dependence of $n_\pm$),
and since $\Gamma \gg 1$ holds,
all the positrons (or electrons) migrate with velocity $+c$ (or $-c$).
As a result, particles at $\xi=\xi^{k-1}$ migrate to $\xi=\xi^k$
after time $(\xi^k-\xi^{k-1})/c$.
Figure~\ref{fig:char} shows this situation,
describing the particle propagation along the characteristics,
whose gradient is given by
(see eq.~\ref{eq:boltz_2})
\begin{equation}
  \frac{d\Gamma}{d\xi}
  = \Ell -\frac{\Pcv(\xi,\Gamma)}{m_{\rm e}c^2\omgp}
  \label{eq:char-eq}
\end{equation}
in the phase space ($\xi$,$\Gamma$).
Adding the produced pairs between $\xi^{k-1}$ and $\xi^k$,
which are depicted by open circles in figure~\ref{fig:char},
we obtain
\begin{equation}
  \sum_l n_+{}_{,l}^k 
  = \sum_l n_+{}_{,l}^{k-1} +D^k \sum_l S_+{}_l^k
  \label{eq:sum_n+}
\end{equation}
for positrons, where $D^k \equiv \xi^k-\xi^{k-1}$.
Equation~(\ref{eq:sum_n+}) holds even if IC scatterings,
which do not change the total particle number, contribute.
In the same manner, we obtain 
\begin{equation}
  \sum_l n_-{}_{,l}^k 
  = \sum_l n_-{}_{,l}^{k-1} -D^k \sum_l S_-{}_l^k
  \label{eq:sum_n-}
\end{equation}
for electrons.
Adding both sides of these two equations,
and utilizing equation~(\ref{eq:sum_src}),
we obtain
\begin{equation}
  \sum_l (n_+{}_{,l}^k +n_-{}_{,l}^k)
  = \sum_l (n_+{}_{,l}^{k-1} +n_-{}_{,l}^{k-1}).
  \label{eq:sum_nn}
\end{equation}

On these grounds, we can conclude that the current density
\begin{equation}
  j_{\rm tot}^k \equiv \sum_l (n_+{}_{,l}^k+n_-{}_{,l}^k)
  \label{eq:consv}
\end{equation}
is virtually kept constant for $k$ (i.e., for position $\xi=\xi^k$),
provided $\Gamma \gg 1$.

We solve the advection-type differential equations~
(\ref{eq:boltz_2}) and (\ref{eq:boltz_3})
on the two-dimensional phase space ($\xi$, $\Gamma$)
by the Cubic Interpolated Propagation (CIP) method
(e.g., Yabe \& Aoki 1991, Yabe, Xiao, \& Utsumi 2001).

%

\subsection{Gamma-ray Boltzmann Equations}
\label{sec:boltz_gamma}
In this subsection, we consider the $\gamma$-ray Boltzmann equations.
For simplicity, we assume 
that the 
density of outwardly (or inwardly) propagating $\gamma$-rays
decreases (or increases) at the same rate with the magnetic field
strength;
this assumption gives a good estimate 
when $W \ll \rlc$ holds.
Under this assumption, the $\gamma$-ray Boltzmann equations become
(e.g., Paper~VII)
\begin{eqnarray}   
  \lefteqn{\pm cB \frac{\partial}{\partial s} 
                  \left( \frac{G_\pm}{B}\right)
     = - \int d\Gamma \frac{\partial\etaP}{\partial\Gamma} 
         \cdot G_\pm(s,\Eg)}
  \nonumber \\
  &+& \int d\Gamma \left[ \etaICg(\Eg,\Gamma,\mu_\pm)
                         +\etaCV( \Eg,\Gamma)
                   \right]
                   N_\pm(s,\Gamma),
  \nonumber \\
  \label{eq:Boltz_gam_0}
\end{eqnarray}   
where $\etaCV$ is the curvature radiation rate [s${}^{-1}$]
into the energy interval $m_{\rm e}c^2(\Eg+d\Eg)$
by a particle migrating with Lorentz factor $\Gamma$.
The first term in the right-hand side expresses the $\gamma$-ray
absorption rate by pair production,
while the second term does the emission rate via
IC scatterings and curvature radiation.

Integrating equations~(\ref{eq:Boltz_gam_0}) over $\Eg$ between
$b_{i-1}$ and $b_i$, 
we obtain
\begin{eqnarray}
  \lefteqn{ \left( \frac{d g_\pm}{d\xi} \right)_i
            = \frac{d(\ln B)}{d\xi} g_\pm{}_{,i}
              \mp \sum_l \etaP{}_{\pm,l}(b_i) \cdot g_\pm{}_{,i}}
  \nonumber \\
  &\pm& \sum_l \left[ {\etaICg}_{\pm,i}(l)+{\etaCV}_{\pm,i}(l)
             \right]
             n_\pm(\xi,b_l),
  \label{eq:boltz_gam_1}
\end{eqnarray}
where $\etaCV{}_+{}_{,i}(l)$ refers to the curvature
radiation rate $\etaCV(b_i,b_l)$ by a positron having Lorentz factor
$\Gamma=b_l$.
In the same manner, $\etaCV{}_-{}_{,i}(l)$ does that by an electron.
Evaluating the right-hand side at $\xi=\xi^{k-1}$,
we obtain the advanced distribution functions
by the Euler method as follows:
\begin{equation}
  g_\pm{}_{,i}^k
  = g_\pm{}_{,i}^{k-1}
   +D^k \left( \frac{d g_\pm}{d\xi} \right)_i^{k-1}.
  \label{eq:boltz_gam_2}
\end{equation}

\subsection{Boundary Conditions}
\label{sec:BD}
As we have described, the set of Maxwell and Botzmann equations 
consist of equations~(\ref{eq:basic-1}), (\ref{eq:basic-2}),
(\ref{eq:boltz_2}), (\ref{eq:boltz_3}), 
and (\ref{eq:boltz_gam_2}).
The Poisson equation and the $\gamma$-ray Boltzmann equations
are ordinary differential equations,
which can be straightforwardly solved by a simple discretization.
On the other hand, the hyperbolic-type partial differential 
equations~(\ref{eq:boltz_2}) and (\ref{eq:boltz_3})
are solved by the CIP method. 

In this section, we consider the boundary conditions 
to solve the set of Maxwell and Boltzmann equations.
At the {\it inner} (starward) boundary
($\xi= \xi^{\rm in}$), we impose (Paper~VII)
\begin{equation}
  \Ell(\xi^{\rm in})=0,
  \quad
  \psi(\xi^{\rm in}) = 0,
  \label{eq:BD-1}
\end{equation}
\begin{equation}
  g_+^i(\xi^{\rm in})=0  
  \quad (i=l_0+1,\ldots,-1,0,1,2,\ldots,l_m),
  \label{eq:BD-3}
\end{equation}
\begin{equation}
  n_+{}_{,l}(\xi^{\rm in})
  = \left\{
      \begin{array}{rl}
        j^{\rm in}, & \quad \mbox{for $l=1$} \\
        0         , & \quad \mbox{for $l=2,3,\ldots,l_m$}
      \end{array} \right.
  \label{eq:BD-4}
\end{equation}
where the dimensionless positronic 
injection current across the inner boundary $\xi=\xi^{\rm in}$ 
is denoted as $j^{\rm in}$.
As we described at the end of \S~\ref{sec:boltz_part},
current density is conserved along the magnetic flux tube
if particles are accelerated well above the initial Lorentz factor.
Thus, we obtain at $\xi=\xi^{\rm in}$
\begin{equation}
  \sum_l n_-{}_{,l}(\xi^{\rm in})= j_{\rm tot}-j^{\rm in}.
  \label{BD-5}
\end{equation}

At the {\it outer} boundary ($\xi=\xi^{\rm out}$), we impose
\begin{equation}
  \Ell(\xi^{\rm out})=0,
  \label{eq:BD-6}
\end{equation}
\begin{equation}
  g_-^i(\xi^{\rm out})=0 \quad 
  \quad (i=l_0+1,\ldots,-1,0,1,2,\ldots,l_m),
  \label{eq:BD-7}
\end{equation}
\begin{equation}
  n_-{}_{,l}(\xi^{\rm out})
  = \left\{
      \begin{array}{rl}
        j^{\rm out}, & \quad \mbox{for $l=1$} \\
        0          , & \quad \mbox{for $l=2,3,\ldots,l_m$}
      \end{array} \right.
  \label{eq:BD-8}
\end{equation}
The current density created in the gap per unit flux tube
can be expressed as
\begin{equation}
  j_{\rm gap}= j_{\rm tot} -j^{\rm in} -j^{\rm out}.
  \label{eq:Jgap}
\end{equation}
We adopt $j_{\rm gap}$, $j^{\rm in}$, and $j^{\rm out}$
as the free parameters.

We have totally $2l_m+2(l_m-l_0+1)+4$ boundary conditions 
(\ref{eq:BD-1})--(\ref{eq:BD-8})
for $2l_m+2(l_m-l_0+1)+2$ unknown functions
$n_\pm{}_l$, $g_\pm{}_i$, $\Psi$, and $\Ell$.
Thus two extra boundary conditions must be compensated 
by making the positions of the boundaries 
$\xi^{\rm in}$ and $\xi^{\rm out}$ be free.
The two free boundaries appear because $\Ell=0$ is imposed at 
{\it both} the boundaries and because $j_{\rm gap}$ is externally imposed.
In other words, the gap boundaries 
($\xi^{\rm in}$ and $\xi^{\rm out}$) shift,
if $j^{\rm in}$ and/or $j^{\rm out}$ varies.

In practice, we give $n_+(\xi^{\rm in},\Gamma)$ and 
$n_-(\xi^{\rm in},\Gamma)$ by a continuous function of $\Gamma$
near the lowest $\Gamma$ bin, $b_1$,
to avoid numerical oscillation due to the
step function represented by equations~(\ref{eq:BD-4}) 
and (\ref{eq:BD-8}).
Nevertheless, detailed functional forms of
the injected particle spectrum little affect the results.

\subsection{Gap Position vs. Particle Injection}
\label{sec:position}
Let us briefly examine how the gap position changes
as a function of $j^{\rm in}$ and $j^{\rm out}$.
For a detailed argument, see \S~2.4 in Paper~VII.
Defining the particle spatial number density by
\begin{equation}
  \bar{n}_\pm \equiv \int_1^\infty d\Gamma n_\pm(\xi,\Gamma),
  \label{eq:sp_n}
\end{equation}
we can rewrite the Poisson equation~(\ref{eq:basic-2}) as
\begin{equation}
  \frac{d\Ell}{d\xi}
  = \frac{B(\xi)}{B^{\rm in}} 
    \left[ n_{\rm gap}(\xi)+f_{\rm null}(\xi)
    \right],
  \label{eq:Poisson_1Db}
\end{equation}
where the created charge density in the gap is defined as
\begin{equation}
  e n_{\rm gap}(\xi)
  \equiv e \left\{  \bar{n}_+(\xi)-j^{\rm in}
                  -[\bar{n}_-(\xi)-j^{\rm out}]
           \right\}
  \label{eq:def_ngap}
\end{equation}
and 
\begin{equation}
  f_{\rm null}(\xi)
  \equiv j^{\rm in} -j^{\rm out} +\frac{B^z(\xi)}{B(\xi)};
  \label{eq:def_fnull}
\end{equation}
the two-dimensional screening effect 
(i.e., the $-\psi/\Delta_\perp^2$ term) is neglected 
in equation~(\ref{eq:Poisson_1Db}).
The created charge density increases outward due to discharge
and satisfies
$ n_{\rm gap}(\xi^{\rm in})  = -j_{\rm gap}$ and 
$ n_{\rm gap}(\xi^{\rm out}) =  j_{\rm gap}$.
We have $n_{\rm gap}<0$ and $f_{\rm null}>0$ 
in the inner part of the gap,
while we have $n_{\rm gap}>0$ and $f_{\rm null}<0$ 
in the outer part.
It follows from equation~(\ref{eq:Poisson_1Db}) that the 
created particles partially screen the original electric field.
If $j_{\rm gap}=j^{\rm in}-j^{\rm out}+B^z(\xi^{\rm in})/B^{\rm in}$
holds,
we obtain $d\Ell/d\xi=0$ at the inner boundary;
that is, the gap is quenched due to the overfilling of space charges.

Then, how about the injected particles?
Let us neglect the created particles by imposing $j_{\rm gap} \ll 1$,
and consider the following three cases:\\
$\bullet$ \ $j^{\rm in}=j^{\rm out}$ \\
In this case, $\Ell$ maximizes (i.e., $f_{\rm null}$ vanishes)
at the point where $B^z=0$ holds.
That is, the gap is located around the null surface.\\
$\bullet$ \ $j^{\rm in} \sim 1$, $j^{\rm out} \sim 0$\\
In this case $\Ell$ maximizes at the point where $B^z=-B$ holds.
That is, the gap is located near to the light cylinder.\\
$\bullet$ \ $j^{\rm in} \sim 0$, $j^{\rm out} \sim 1$\\
In this case $\Ell$ maximizes at the point where $B^z=B$ holds.
That is, the gap is located near to the polar cap surface.\\

In short, created particles in the gap can quench the gap,
while injected particles consisting of a single charge only shift it.
The gap center is located at the \lq generalized null surface'
on which $f_{\rm null}$ vanishes,
provided that the created particle density is negligible 
(i.e., if $n_{\rm gap} \ll 1$).
Strictly speaking, injected particles emit $\gamma$-rays
that can materialize as pairs;
therefore, injected particles contribute for screening of $\Ell$
in a sense.
Nevertheless, $\Ell$ is screened by the created particles
{\it discharged} by $\Ell$ itself, 
not by the originally injected ones that {\it do not return}.

\section{Application to Individual Pulsars}
\label{sec:app}
To solve the set of the Maxwell and Boltzmann equations,
we must specify the
X-ray field, $dF_{\rm s}/d\Es$, which is necessary to
compute the pair-production redistribution function
(eq.~[\ref{eq:def_etaP_2}]).
In this paper, we use the X-ray fluxes and spectra
observed for individual rotation-powered pulsars
to specify $dF_{\rm s}/d\Es$.
In \S~\ref{sec:app_1}, we summarize the observed properties
of the X-ray field.
Then we apply the theory to
the Vela pulsar in \S~\ref{sec:app_vela},
to PSR~B1706-44 in \S~\ref{sec:app_1706},
to the Geminga pulsar in \S~\ref{sec:app_Gemi},
and to PSR~B1055-52 in \S~\ref{sec:app_1055}.
We assume that the solid angle of the emitted $\gamma$-rays is 1~ster
throughout this paper.

\subsection{Input Soft Photon field}
\label{sec:app_1}
We consider the photons emitted from the
neutron star surface as the seed photons for
($\gamma$-$\gamma$) pair-production and IC scatterings.
That is, we do not consider power-law X-ray components,
because they are probably magnetospheric and beamed away from 
the accelerator.
We evaluate the IR photon field, which is needed to compute the
IC scattering rate, from the Rayleigh-Jeans tail of the surface
(blackbody) component.
In table~1, we present the observed properties of the 
four $\gamma$-ray pulsars
exhibiting surface X-ray components,
in order of spin-down luminosity, $L_{\rm spin}$.

\begin{table*}
  \centering
    \begin{minipage}{160mm}
      \caption{Input thermal X-ray field}
      \begin{tabular}{@{}lccccrcrcll@{}}
        \hline
        \hline
        pulsar	
		& $\lg L_{\rm spin}$
		& distance
		& $\Omega$		& $\lg B_{\rm s}$
		& $kT_{\rm s}$		& $A_{\rm s}/A_*$
		& $kT_{\rm h}$		& $A_{\rm h}/A_*$
		& model
		& refs.				\\
        \	
		& ergs s${}^{-1}$
		& kpc
		& rad s${}^{-1}$	& G
		& eV			&
		& eV			& 
		& 
		&					\\
        \hline
        Vela	
		& 36.84		
		& 0.25
		& 70.4		& 12.53
		& 128		& 0.045
		& $\ldots$	& $\ldots$
 		& blackbody
		& 1,2 				\\
        \ 	
		& \ 
		& \ 
		& \ 		& \ 
		& 59		& 1.000
		& $\ldots$	& $\ldots$
		& hydrogen atm.
		& 1 				\\
        B1706-44
		& 36.53
		& 2.50
		& 61.3		& 12.49
		& 143		& 0.129
		& $\ldots$	& $\ldots$
 		& blackbody
		& 3					\\
        Geminga
		& 34.51		
		& 0.16
		& 26.5		& 12.21
		& 48		& 0.160
		& $\ldots$	& $\ldots$
 		& blackbody
		& 4, 5				\\
        B1055--52
		& 34.48		
		& 1.53
		& 31.9		& 12.03
		& 68		& 7.300
		& 320		& $10^{-3.64}$
 		& blackbody
		& 6					\\
        \hline
      \end{tabular}
      \begin{flushleft}
    1: Pavlov et al. 2001;                  
    \quad
    2: $\ddot{\rm O}$gelman et al. 1993;   
    \quad
    3: Gotthelf, Halpern, Dodson 2002;      
    \quad
    4:  Halpern \& Wang 1997;                
    \quad
    5:  Becker \& Tr$\ddot{\rm u}$mper 1996; 
    \quad
    6: Greiveldinger et al. 1996;           
    \quad
      \end{flushleft}
    \end{minipage}
\end{table*}

{\bf Vela} (J0835--4513)\ 
From Chandra observations in 0.25-8.0~keV,
the spectrum of this pulsar is turned out to consist of
two distinct component:
A soft component and a hard, power-law component.
The hard component is probably a magnetospheric origin
and hence will be beamed away from the gap.
Thus, we consider only the soft component
as the X-ray field illuminating the outer gap.
This component can be modeled 
either as a blackbody spectrum with $kT=1.49$MK and
$A_{\rm s}=0.045 A_\ast(d/0.25)^2$,
or as a magnetic hydrogen atmosphere spectrum
with effective temperature $kT=0.68$MK
(Pavlov et al.~2001). 
Based on high-resolution Ca~{\small II} and Na~{\small I}
absorption-line spectra toward 68 OB stars in the direction
of the Vela supernova remnant,
Cha, Sembach, and Danks (1999) determined the distance to be
$250 \pm 30$~pc.
{\bf B1706--44} (J1710--4432)\ 
Gotthelf, Halpern, Dodson (2002) reported
a broad, single-peaked pulsed profile with pulsed fraction of $23 \%$,
using the High Resolution Camera on-board the Chandra X-ray observatory.
They fitted the spectroscopic data to find
(at least) two components, 
e.g., a blackbody of $kT=143$~eV with $A= 0.129 A_\ast$
and a power-law component with photon index of $-2.0$.
We consider that the former component illuminates the gap efficiently
and neglect the second one.
We adopt $d=2.5$~kpc as a compromise between the smaller 
dispersion-measure distance of $1.8$~kpc based on the
free electron model by Taylor and Cordes (1993)
and the larger H~I kinematic distance of $2.4$--$3.2$~kpc
derived by Koribalski et al.~(1995).\\
{\bf Geminga} (J0633+1746)\ 
The X-ray spectrum consists of two components:
the soft surface blackbody with $kT_{\rm s}=50$ eV and 
$A_{\rm s}= 0.22 A_* (d/0.16)^2$
and a hard power law with $\alpha= -1.6$
(Halpern \& Wang 1997). 
A parallax distance of 160pc was estimated from HST observations
(Caraveo et al. 1996). \\
{\bf B1055--52} (J1059--5237)\ 
Combining ROSAT and ASCA data, Greiveldinger et al. (1996)
reported that the X-ray spectrum consists of two components:
a soft blackbody with $kT_{\rm s}=68$ eV and 
$A_{\rm s}= 7.3 A_* (d/1.53)^2$
and a hard blackbody with $kT_{\rm h}=320$ eV and 
$A_{\rm h}= 2.3 \times 10^{-4} (d/1.53)^2$.
The distance is estimated to be $1.53$~kpc from
dispersion measure (Taylor \& Cordes 1993).\\

\subsection{The Vela pulsar}
\label{sec:app_vela}
We first apply the theory to the Vela pulsar,
using the X-ray field modeled by the blackbody spectrum 
in \S~\ref{sec:Ell_char}-\ref{sec:powerlaw},
and by the hydrogen atmosphere spectrum in \S~\ref{sec:hydro}.

\subsubsection{Acceleration Field and Characteristics}
\label{sec:Ell_char}
Let us first consider the spatial distribution of the acceleration 
field,
\begin{equation}
  -\frac{d\Psi}{ds}
    = \frac{\omgp}{c} \frac{m_{\rm e}c^2}{e}
      \Ell(\xi) \quad [\mbox{V m}^{-1}],
\end{equation}
which is solved from the Poisson equation~(\ref{eq:basic-2}).
We may notice here that $s$ is related with $\xi$ 
by equation~(\ref{eq:def-xi}).
For the Vela pulsar, a small value 
$j_{\rm gap}=4.6 \times 10^{-5}$ gives the best-fit spectrum
(see \S~\ref{sec:powerlaw} for details). 

\begin{figure} 
\centerline{ \epsfxsize=8.5cm \epsfbox[200 20 470 200]
              {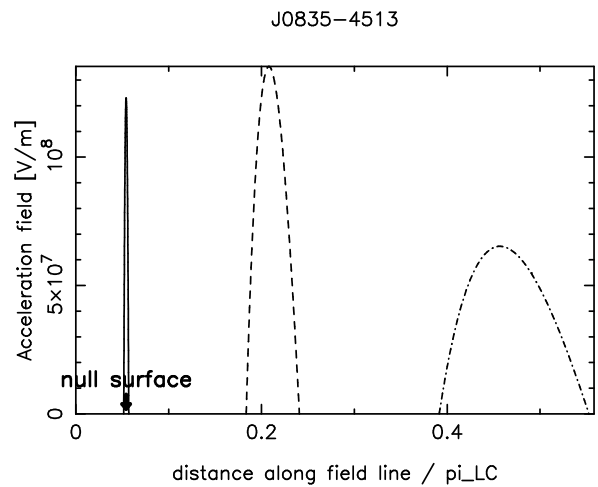} } 
\caption{
Spatial distribution of $-d\Psi/ds$ 
for $j^{\rm in}=0$ (solid), $0.25$ (dashed), 
and $0.5$ (dash-dotted),
for the Vela pulsar when $\inc= 75^\circ$
and $j_{\rm gap}=4.6 \times 10^{-5}$ and $j^{\rm out}=0$.
The abscissa designates the distance along the last-open field line
normalized by the light cylinder radius.
        }
\label{fig:EVela_75z} 
\end{figure}

To compare the effects of particle injection,
we present the $\Ell$ distribution
for the three cases of 
$j^{\rm in}=0$ (solid), 0.25 (dashed), and 0.50 (dash-dotted)
in figure~\ref{fig:EVela_75z}.
The magnetic inclination is chosen to be $\inc=75^\circ$.
We adopt $j^{\rm out}=0$ throughout this paper,
unless its value is explicitly specified.
The abscissa denotes the distance along the last-open field line
(i.e., the $s$ coordinate used 
 in eqs.~[\ref{eq:def_rhoGJ}] and [\ref{eq:Poisson_2D}])
normalized by $\rlc$.

As the solid line shows, $\Ell$ is located around the null surface
when there is no particle injection across either of the boundaries.
Moreover, $\Ell$ varies quadratically, because the Goldreich-Julian
charge density deviates from zero nearly linearly near to the
null surface.

As the dashed and dash-dotted lines indicate,
the gap shifts outwards as $j^{\rm in}$ increases.
When $j^{\rm in}=0.5$ for instance, 
the gap is located on the half way
between the null surface and the light cylinder.
This result is consistent with Papers~VII, VIII, and IX.

\begin{figure} 
\centerline{ \epsfxsize=8.5cm \epsfbox[200 20 470 200]
              {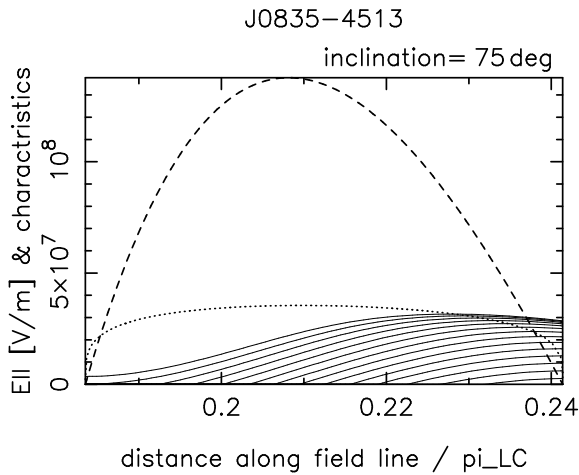} } 
\caption{
Spatial distribution of $-d\Psi/ds(s)$ (dashed)
for the Vela pulsar when $\inc= 75^\circ$,
$j^{\rm in}=0.25$, $j^{\rm out}=0$,
and $j_{\rm gap}= 4.6 \times 10^{-5}$.
The soft photon field is approximated by a single blackbody 
component.
The characteristics are also shown by solid lines.
The equilibrium Lorentz factor, which would be obtained
if we balanced the curvature radiation-reaction force
and the electrostatic force, 
is indicated by the dotted line.
        }
\label{fig:EVela_75a} 
\end{figure}

In figure~\ref{fig:EVela_75a},
we present the characteristics of 
partial differential equation~(\ref{eq:boltz_2})
for positrons by solid lines, together with 
$\Ell(\xi)$ when $j^{\rm in}=0.25$ and $\inc=75^\circ$
(i.e., the dashed line in fig.~\ref{fig:EVela_75z}).
We also superpose the equilibrium Lorentz factor 
that would be obtained
if we assumed the balance between the curvature radiation reaction
and the electrostatic acceleration, as the dotted line.
It follows that the particles are not saturated at the
equilibrium Lorentz factor in most portions of the gap.

In the outer part of the gap where $\Ell$ is decreasing,
characteristics begin to concentrate; as a result, 
the energy distribution of outwardly propagating particles
forms a \lq shock' in the Lorentz factor direction.
However, the particle Lorentz factors do not match 
the equilibrium value (dotted line).
For example, near the outer boundary,
the particles have larger Lorentz factors compared with the
equilibrium value, 
because the curvature cooling scale is longer than the gap width.
Thus, we must discard the mono-energetic approximation 
that all the particles migrate at the equilibrium Lorentz factor
as adopted in Papers~I through IX.
We instead have to solve the energy dependence of the particle 
distribution functions explicitly.

The particles emit $\gamma$-rays not only inside of the gap
but also outside of it,
being decelerated by the curvature radiation-reaction force.
The length scale of the deceleration is given by
\begin{eqnarray}
  l_{\rm curv} 
  &=& c \cdot \frac{\Gamma m_{\rm e}c^2}
                   {\displaystyle\frac{2e^2}{3c^3}\Gamma^4
                           \left(\frac{c^2}{\rho_{\rm c}}\right)^2}
  \nonumber\\
  &=& 0.4 \rlc \Omega_2{}^{-1}
      \left(\frac{\Gamma}{10^7}\right)^{-3}
      \left(\frac{\rho_{\rm c}}{\rlc/2}\right)^2
  \label{eq:cool_curv}
\end{eqnarray}
Since the typical Lorentz factor is a few times of $10^7$,
$l_{\rm curv}$ is typically much less than $\rlc$.
Therefore, the escaping particles lose their energy well inside 
of the light cylinder.


\subsubsection{Particle Energy Distribution}
\label{sec:spc_particle}

\begin{figure} 
\centerline{ \epsfxsize=8.5cm \epsfbox[200 20 470 200]
              {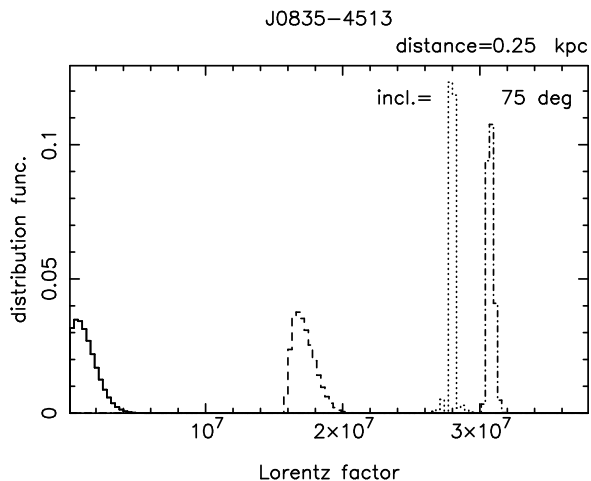} } 
\caption{
Particle energy distribution at several points along the magnetic field
lines for the same case as in figure~\ref{fig:EVela_75a}.
Initial spectrum (solid line) 
evolves to dashed, dash-dotted, and dotted lines,
as particles propagate outwards. 
}
\label{fig:PVela_75a} 
\end{figure}

As we have seen in the foregoing subsection,
the distribution function of the outwardly propagating particles
forms a \lq shock' in the outer part of the gap.
In figure~\ref{fig:PVela_75a},
we present the energy distribution of particles at several 
representative points along the field line.
At the inner boundary ($s=0.184 \rlc$),
particles are injected with Lorentz factors
typically less than $4\times 10^6$ 
as indicated by the solid line.
Particles migrate along the characteristics in the phase space
and gradually form a \lq shock' 
as the dashed line (at $s=0.205\rlc$) indicates,
and attains maximum Lorentz factor 
at $s=0.228\rlc$ as the dash-dotted line indicates.
Then they begin to be decelerated gradually
and escape from the gap
with large Lorentz factors $\sim 2.5 \times 10^7$ (dotted line)
at the outer boundary, $s=s^{\rm out}=0.241\rlc$.

Even though the \lq shock' is captured by only a few  
grid points for the dash-dotted line in figure~\ref{fig:PVela_75a},
the CIP scheme accurately satisfies 
the conservation of the total current density,
\begin{eqnarray}
  j_{\rm tot}
  &=& n_+(\xi)+n_-(\xi)
  \nonumber\\
  &=& j^{\rm in}+j^{\rm out}+j_{\rm gap} \approx j^{\rm in}.
  \label{eq:consv2}
\end{eqnarray}
For this case,
$j_{\rm tot}$ is accurately conserved at $0.25$ 
within 0.2\% errors even at the \lq shock'.

\subsubsection{Formation of Power-law Gamma-ray Spectrum}
\label{sec:powerlaw}
So far, we have seen that the outwardly propagating particles 
are not saturated at the equilibrium value
and that such particles escape from the gap
with sufficient Lorentz factors suffering subsequent cooling 
via curvature process.
It seems, therefore, reasonable to suppose that
a significant fraction of the $\gamma$-ray luminosity 
is emitted from such escaping particles.

\begin{figure} 
\centerline{ \epsfxsize=8.5cm \epsfbox[200 20 470 200]
              {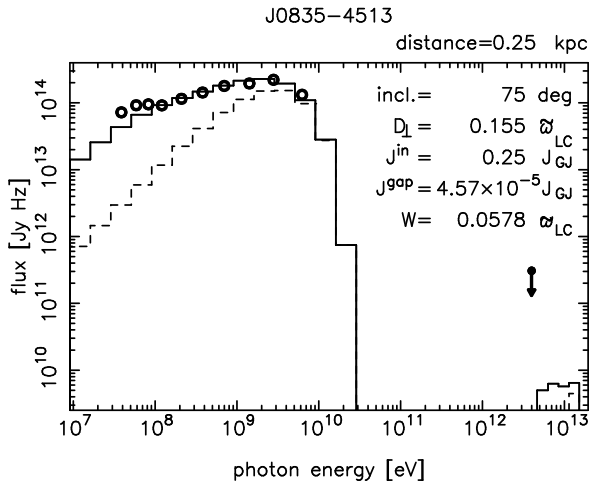} } 
\caption{
Computed $\gamma$-ray spectrum for the Vela pulsar
for the same case as in figure~\ref{fig:EVela_75a}.
The dashed line represents the emission from the gap,
while the solid one includes the flux emitted by the particles
being decelerated via curvature process outside of the gap.
The soft photon field is modeled as a blackbody.
        }
\label{fig:SVela_75a}
\end{figure}

We present the $\gamma$-ray spectrum 
emitted from outwardly propagating particles (i.e., positrons)
for the Vela pulsar with $\inc=75^\circ$,
$j_{\rm gap}= 4.6 \times 10^{-5}$, 
$j^{\rm in}= 0.25$ in figure~\ref{fig:SVela_75a}.
The dashed line represents the $\gamma$-ray flux emitted
within the gap,
while the solid one includes that emitted outside of the gap
by the escaping particles.
Therefore, the difference between the solid and the dashed
lines indicates the $\gamma$-ray flux emitted 
by the particles migrating outside of the gap.
For comparison, we plot the phase-averaged 
EGRET spectrum, which is approximated by a
power law with a photon index $-1.7$ (Kanbach et al. 1994).

\begin{figure} 
\centerline{ \epsfxsize=8.5cm \epsfbox[200 20 470 200]
              {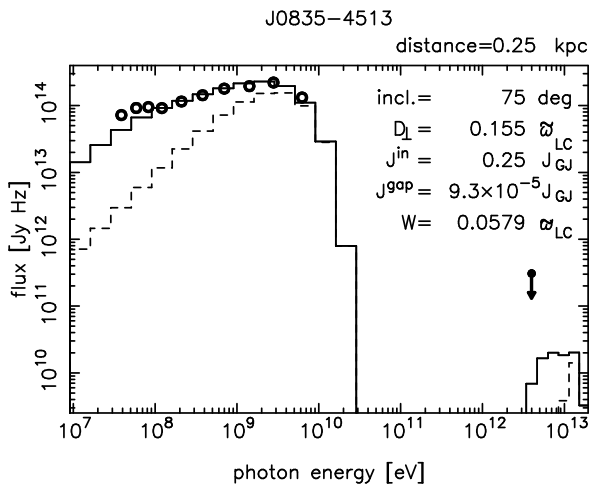} } 
\caption{
Same figure as fig~\ref{fig:SVela_75a},
except that the soft photon field is modeled as 
a hydrogen atmosphere spectrum.
       }
\label{fig:SVela_75b}
\end{figure} 

It follows from the figure that the $\gamma$-ray spectrum 
in 100~MeV--GeV energies can be explained by the 
curvature radiation emitted by the escaping particles.
We adjusted the transfield thickness as
$D_\perp=0.16\rlc=2.8W$ so that the observed flux may be explained.
The luminosity of the $\gamma$-rays emitted outside of the gap
contribute 48\% of the total luminosity 
$5.08 \times 10^{33} \mbox{ergs s}^{-1}$
between 100~MeV and 20~GeV.
In another word, we do not have to assume a power-law energy
distribution for particles 
(as assumed in some of the previous outer-gap models)
to explain the power-law $\gamma$-ray spectrum for the Vela pulsar.
This conclusion is natural, 
because a power-law energy distribution of particles will not be 
achieved by an electrostatic acceleration,
and because magnetohydrodynamic shocks (i.e., real shocks)
will not be formed in the accelerator.

Because the X-ray field is dense for this young pulsar,
the pair-production mean free path, 
and hence the gap width becomes small
(for details, see 
 Hirotani \& Okamoto 1998, 
 Hirotani 2000a, Paper~IV;
 Hirotani 2001,  Paper~V).
As a result, the potential drop in the gap 
$2.24 \times 10^{13}$~V is only 0.81~\% of the 
electro-motive force (EMF) exerted on the spinning neutron star surface
$\sim \mu/\rlc^2 = 2.79 \times 10^{15}$~V.
Nevertheless, this potential drop is enough
to accelerate particles into high Lorentz factors, $10^{7.3}$.

\subsubsection{Hydrogen Atmosphere Model}
\label{sec:hydro}
We next briefly examine the $\gamma$-ray spectrum when the
soft photon field is modeled as the hydrogen atmosphere spectrum
(Pavlov et al.~2001), rather than the blackbody spectrum.
In figure~\ref{fig:SVela_75b},
we present the resultant spectrum for the same parameter set
as in figure~\ref{fig:SVela_75a},
except for the fitting model for the observed X-ray field.
We fix $s^{\rm in}$ instead of $j_{\rm gap}$ for comparison purpose.
It follows that IC flux increases 3 times while 
$j_{\rm gap}$ does only 2 times.
This is because the hydrogen atmosphere model gives much low-energy
(IR) photons compared with the blackbody model, 
as can be understood from its small effective temperature and large
emitting area.
Nevertheless, the obtained GeV spectrum little differ 
from the blackbody model.
This is due to the \lq negative feedback effect'
of the gap electrodynamics (Paper~V),
which ensures the existence of solutions in a wide parameter range
and suggests the dynamical stability of the solutions.

\subsubsection{Solutions in a Wide Parameter Space}
\label{sec:Vela_param}
With both blackbody and the hydrogen atmosphere models,
we can explain the observed $\gamma$-ray spectrum 
in a wide parameter space
$45^\circ < \inc < 75^\circ$ and 
$0.125 < j^{\rm in} < 0.25$,
by appropriately choosing $j_{\rm gap}$ and $D_\perp$.
With increasing $j_{\rm gap}$ ($\ll 1$),
the $\nu F_\nu$ [Jy~Hz] peak energy increases 
because of the increased $W$, 
while the sub-GeV spectrum becomes hard
because of the significant $\gamma$-ray emission within the extended gap 
(rather than outside of it) above GeV energies.
On the other hand, $D_\perp$ affects only 
the normalization of the $\gamma$-ray flux;
the $\gamma$-ray luminosity is proportional to $D_\perp^2$.

Let us first fix $j^{\rm in}$ at $0.25$ and consider 
how the best-fit values of $j_{\rm gap}$ and $D_\perp$
depend on $\inc$.
As we have seen, they are $j_{\rm gap}=4.6 \times 10^{-5}$ and 
$D_\perp=0.155\rlc=0.84 s^{\rm in}$ for $\inc=75^\circ$, 
where the gap inner boundary is located at
$s=s^{\rm in}=0.1835\rlc$.
However, the ratio $D_\perp / s^{\rm in}$ 
increases with decreasing $\inc$ and becomes $1.06$ for $45^\circ$
(cf. $0.84$ for $\inc=75^\circ$).
From geometrical consideration, 
we conjecture that $D_\perp$ should not greatly exceed $s^{\rm in}$;
we thus consider that $\inc > 45^\circ$ is appropriate 
for $j^{\rm in}=0.25$.
A large $\inc$ is preferable to obtain a small $D_\perp / \rlc$.
However, since the radio pulsation shows a single peak, 
we consider that $\inc$ is not too close to $90^\circ$.
On these grounds,
we adopted $\inc=75^\circ$ as a moderate value in this section.

Let us next fix $\inc$ at $75^\circ$ and 
consider how the best-fit values of $j_{\rm gap}$ and $D_\perp$
depend on $j^{\rm in}$. 
The ratio $D_\perp / \rlc$ increases with decreasing $j^{\rm in}$
and becomes $0.94$ for $j^{\rm in}=0.125$
(c.f. $0.84$ for $j^{\rm in}=0.25$).
This is because the decreased flux of the outwardly migrating particles
(due to decreased $j^{\rm in}$) must be compensated by a large $D_\perp$
to produce the same $\gamma$-ray flux.
Thus, we consider $j^{\rm in}>0.125$ is appropriate for $\inc=75^\circ$.
A large $j^{\rm in}$ is preferable to obtain a small $D_\perp / \rlc$.
However, for $j^{\rm in}>0.25$,
the gap is so extended that 
a significant $\gamma$-rays are emitted above GeV within the gap;
as a result, the sub-GeV spectrum becomes too hard.
On these grounds, we adopted $j^{\rm in}=0.25$ as a moderate value.


For $45^\circ < \inc < 75^\circ$,
$0.125 < j^{\rm in}<0.25$,
and appropriately chosen $j_{\rm gap}$ and $D_\perp$,
TeV flux is always less than $3 \times 10^{10}$~JyHz.
Thus, one general point becomes clear:
TeV flux is unobservable with current ground-based telescopes,
provided that the emission solid angle is as large as 1~ster
and that the surface thermal (not magnetospheric) X-rays 
are upscattered inside and outside of the gap.
Since the magnetospheric X-rays will be beamed away from the
gap and their specific intensity is highly uncertain,
we leave the problem of the upscatterings of magnetospheric 
(power-law) X-rays untouched.

\subsection{PSR~B1706-44}
\label{sec:app_1706}
We next apply the theory to a Vela-type pulsar, PSR~B1706-44.
To consider $D_\perp / s^{\rm in}$ as small as possible,
we adopt a large magnetic inclination $75^\circ$.
We compare $\nu F_\nu$ spectra for the three cases:
$j^{\rm in}=0.4$, $0.2$, and $0.1$.

\begin{figure} 
\centerline{ \epsfxsize=8.5cm \epsfbox[200 20 470 200]
              {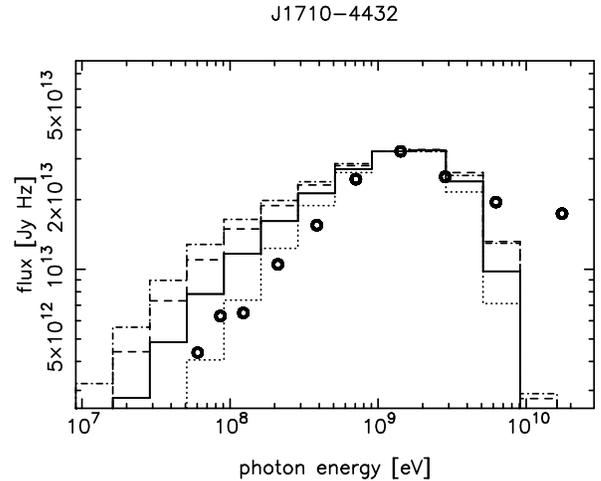} } 
\caption{
Computed $\gamma$-ray spectra for PSR~B1706-44.
The solid, dashed, dash-dotted lines
represent the spectra 
for $j^{\rm in}=0.4$, $0.2$, $0.1$, respectively, 
when $d=2.5$~kpc and $\inc=75^\circ$, 
while the dotted one for 
$j^{\rm in}=0.4$ when $d=1.0$~kpc and $\inc=45^\circ$.
        }
\label{fig:S1706} 
\end{figure}

To examine how the sub-GeV spectrum depends on $j^{\rm in}$, 
we fix the $\nu F_\nu$ peak at the observed value, $\sim 2$~GeV,
by adjusting $j_{\rm gap}$ appropriately.
For the solid, dashed, and dash-dotted lines 
in figure~\ref{fig:S1706}, we adopt
$j_{\rm gap}=2.2 \times 10^{-4}$, $1.8 \times 10^{-4}$, and 
$1.5 \times 10^{-4}$, respectively, and $d=2.5$~kpc.
Moreover, the perpendicular thickness is adjusted so that
the predicted flux may match the observed value 
($3.2 \times 10^{13}$~JyHz) at $1.4$~GeV;
for the solid, dashed, and dash-dotted lines, they are
$D_\perp=1.05\rlc=3.5s^{\rm in}$, 
$0.61\rlc=4.1s^{\rm in}$, $0.49\rlc=5.3s^{\rm in}$, respectively.
The sub-GeV spectrum becomes harder with increasing $j^{\rm in}$,
because the ratio of the flux emitted outside of the gap
and that emitted within it
decreases as the gap extends with increasing $j^{\rm in}$.
However, as the solid line indicates, 
the obtained sub-GeV spectrum is still too soft to match the observation.
For a smaller $j^{\rm in}$ ($<0.1$), 
the sub-GeV spectrum becomes further soft.
For a non-zero $j^{\rm out}$, 
the inwardly shifted gap is shrunk to emit smaller $\gamma$-ray flux, 
which results in a further greater $D_\perp/s^{\rm in}$.
On these grounds, the predicted sub-GeV spectrum becomes too soft 
or the $\gamma$-ray flux becomes too small 
(i.e., $D_\perp$ becomes too large) for any parameter set
of $j^{\rm in}$ and $j^{\rm out}$, if $\inc=75^\circ$ and $d=2.5$~kpc.

For a small inclination ($\inc<75^\circ$),
the gap is located relatively outside of the magnetosphere,
because the null surface crosses the last-open field line
at large distances from the star.
Because the pair-production mean-free path becomes large far from the star, 
the gap is extended for a small magnetic inclination.
In such an extended gap, 
particles saturate at the equilibrium Lorentz factor
at the outer part (Takata et al.~2002)
and emit most of the $\gamma$-rays around the 
central energy of curvature radiation.
As a result, a hard sub-GeV spectrum can be expected;
however, we have to assume a large $D_\perp$
that exceeds $\rlc$ if $d=2.5$~kpc.

Nevertheless,
if $d$ is much less than $2.5$~kpc and $\inc<75^\circ$,
we can explain the observed spectrum with moderate $D_\perp$.
For example,
for $d=1$~kpc, $\inc=45^\circ$, $j^{\rm in}=0.4$, and
$j_{\rm gap}=3.8\times 10^{-3}$,
the gap exists in $0.51\rlc<s<0.74\rlc$
and the resultant spectrum (dotted line in figure~\ref{fig:S1706})
matches the observation relatively well
with a reasonable thickness, $D_\perp=0.80\rlc=1.5 s^{\rm in}$.

On the other hand, for a large inclination ($\inc>75^\circ$), 
the sub-GeV spectrum becomes softer than $\inc=75^\circ$ case
for the same $j^{\rm in}$, $j^{\rm out}$, and $j_{\rm gap}$.
Therefore, the spectrum will not match the observation 
whatever distance we may assume.

We can alternatively consider a CHR-like outer gap.
Assuming a small $D_\perp$ (say, $0.05$),
we find that the gap extends along the field lines 
due to the screening effect of the zero-potential walls.
In most portions of this extended gap,
particles are nearly saturated at the equilibrium Lorentz factor.
As a result, the sub-GeV spectrum becomes hard and match the observation
with appropriate peak energy around $2$~GeV.
However, in this case, the $\gamma$-ray flux becomes too small 
to match the observed value, unless we adopt an unrealistic distance
(say, $300$~pc).

In short, the phase-averaged EGRET spectrum 
for this pulsar cannot be explained by our current model
for any combinations of 
$\inc$, $j^{\rm in}$, $j^{\rm out}$, $j_{\rm gap}$, and $D_\perp$,
if $d=2.5$~kpc. 
Therefore, we suggest a small distance (e.g., $1$~kpc) for this pulsar.

\subsection{The Geminga Pulsar}
\label{sec:app_Gemi}
Let us apply the theory to a cooling neutron star, 
the Geminga pulsar.
For a small $\inc$ (e.g., $45^\circ$), 
the gap is so extended that the outer boundary exceeds the
light cylinder.
For a larger $\inc$, on the other hand,
not only the outer-gap emission, but also a polar-cap one
could be in our line of sight.
Since there has been no radio pulsation confirmed,
we consider a moderate magnetic inclination $\inc=60^\circ$.

\begin{figure} 
\centerline{ \epsfxsize=8.5cm \epsfbox[200 20 470 200]
              {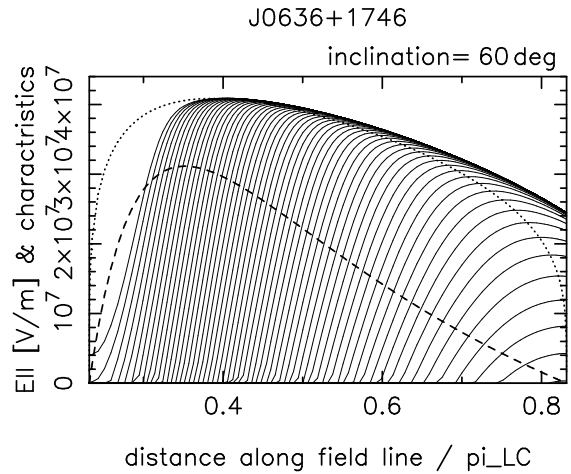} } 
\caption{
Spatial distribution of $-d\Psi/ds$ (dashed line)
for the Geminga pulsar when $\inc= 60^\circ$, $j^{\rm in}= 0.25$,
$j^{\rm out}=0$, and $j_{\rm gap}=8.0 \times 10^{-5}$.
Particles are saturated at the equilibrium Lorentz factor
(dotted line) for this middle-aged pulsar.
        }
\label{fig:EGemi_60} 
\end{figure}

\begin{figure} 
\centerline{ \epsfxsize=8.5cm \epsfbox[200 20 470 200]
              {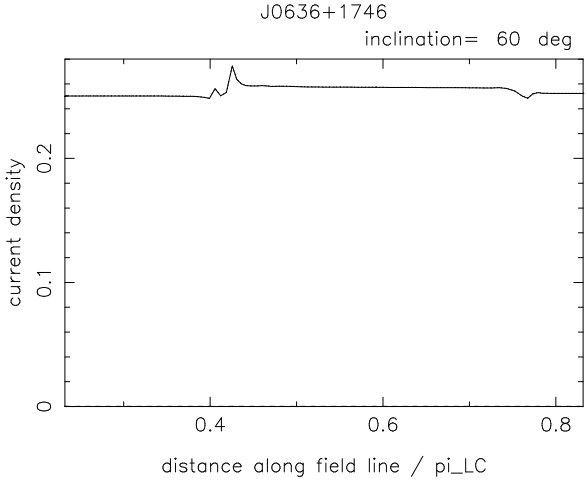} } 
\caption{
Total current density, $j_{\rm tot}$, 
for the same case as in figure~\ref{fig:EGemi_60}.
Even though the particle distribution function forms a strong \lq shock'
in the Lorentz factor direction in $0.4 \rlc < s < 0.45 \rlc$
(see fig.~\ref{fig:EGemi_60}),
$j^{\rm tot}$ is conserved relatively accurately.
        }
\label{fig:CGemi_60} 
\end{figure}

Since the soft photon field is less dense compared with 
young pulsars like Vela or B1706--44,
the gap is extended along the field lines.
For the set of parameters $j^{\rm in}=0.25$, $j^{\rm out}=0$,
and $j_{\rm gap}=8.0\times 10^{-5}$,
which gives the best-fit $\gamma$-ray spectrum,
we obtain $W \sim 0.59 \rlc$.
We present the spatial distribution of $\Ell$ 
obtained for this set of parameters 
as the dashed line in figure~\ref{fig:EGemi_60},
as well as the characteristics (solid lines)
and the equilibrium Lorentz factor (dotted line).
In the outer part of this extended gap,
$\rhoGJ$ gradually increases with the distance, $s$.
As a result, $\Ell(s)$ deviates from quadratic distribution
and decrease gradually in the outer part of the gap.
Because of this extended feature,
particles are nearly saturated at the equilibrium Lorentz factor.
In another word, the mono-energetic approximation adopted in
Papers~I--IX is justified for this middle-aged pulsar.
Particle distribution function forms a strong \lq shock',
which is captured only with one grid point,
in $0.4 < s/\rlc < 0.45$.
As a result,
$j_{\rm tot}$ fluctuates a little 
as figure~\ref{fig:CGemi_60} indicates.
Nevertheless, it returns to the $0.2525$ level,
which is $1$~\% greater than the value it should be ($0.250$),
as the characteristics begin to be less concentrated
beyond $s=0.45 \rlc$.

\begin{figure} 
\centerline{ \epsfxsize=8.5cm \epsfbox[200 20 470 200]
              {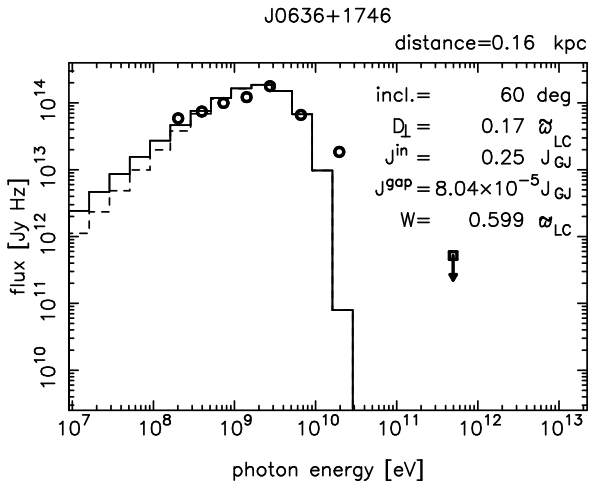} } 
\caption{
Computed $\gamma$-ray spectrum for the Geminga pulsar
for the same case as in figure~\ref{fig:EGemi_60}.
The dashed and solid lines represent the same
components as figure~\ref{fig:SVela_75a}.
        }
\label{fig:SGemi_60} 
\end{figure} 

\begin{figure} 
\centerline{ \epsfxsize=8.5cm \epsfbox[200 20 470 200]
              {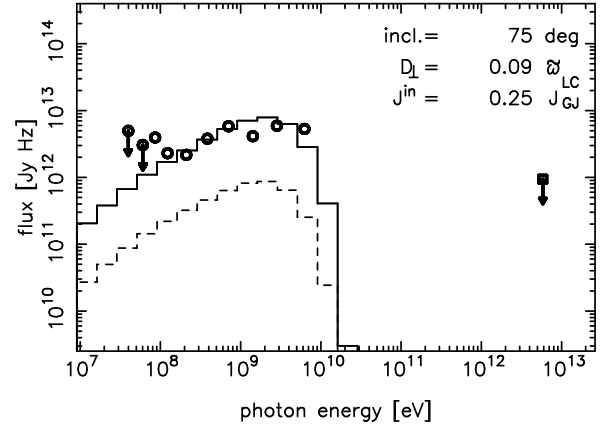} } 
\caption{
Computed $\gamma$-ray spectrum for B1055--52.
Escaping particles little contribute for the luminosity.
The solid and dashed lines correspond to 
$d=0.50$~kpc and $d=1.53$~kpc, respectively.
Parameters are commonly set as
$D_\perp=0.09\rlc$,
$j^{\rm in}= 0.25$, $j^{\rm out}=0$.
        }
\label{fig:S1055_75} 
\end{figure} 

In figure~\ref{fig:SGemi_60}, we present the resultant $\gamma$-ray
spectrum.
Because of the nearly saturated motion of the particles,
they lose most of their energy within the gap.
As a result,
$\gamma$-ray luminosity associated with the escaping particles 
($5.3 \times 10^{31} \mbox{ergs s}^{-1}$), is
negligibly small compare to that emitted within the gap
($1.11 \times 10^{33} \mbox{ergs s}^{-1}$),
which is represented by the dashed line.

It follows from the figure that the observed, soft spectrum
between 200~MeV and 6~GeV can be explained by the
present one-dimensional model.
It is interesting to compare this result
with what obtained for the Vela pulsar 
(figs.~\ref{fig:SVela_75a} and \ref{fig:SVela_75b}).
Between $100$~MeV and $1$~GeV energies,
both spectra are formed by the superposition of the curvature radiation
emitted by the particles having different energies at different
positions.
The important difference is that the particles are saturated at the
equilibrium Lorentz factor {\it in the gap} for the Geminga pulsar,
while they are nearly mono-energetic but only decelerated 
via curvature process {\it outside of the gap} for the Vela pulsar.
Because the particles are no longer accelerated outside of the gap,
they emit $\gamma$-rays in lower energies 
compared with those still being accelerated in the gap.
As a result, the $\gamma$-ray spectrum for the Vela pulsar becomes
softer than that for the Geminga pulsar.
Extending this consideration,
we can predict that a $\gamma$-ray spectrum below GeV
is soft for a young pulsar 
and tends to become hard as the pulsar ages.


\subsection{PSR~B1055-52}
\label{sec:app_1055}
Let us finally apply the present theory to another middle-aged pulsar,
B1055-52.
To obtain a large $\gamma$-ray flux for an appropriately chosen set of
$j^{\rm in}$, $j_{\rm gap}$, and $D_\perp(<s^{\rm in})$,
we adopt a large magnetic inclination, $\inc=75^\circ$.

Since the acceleration field and the particle energy distributions
are similar to the Geminga pulsar, we present only the 
computed $\gamma$-ray spectra for this pulsar 
in figure~\ref{fig:S1055_75}. 
The solid and dashed lines represent the spectra for
the distances $0.50$ and $1.53$~kpc, respectively.
For $d=0.50$~kpc, $j_{\rm gap}= 4.5 \times 10^{-3}$ is chosen
so that the peak energy of curvature radiation
may match the observed peak energy.
In this case, the gap exists in $0.1450\rlc < s < 0.3882\rlc$.
For $d=1.53$~kpc,
$j_{\rm gap}=4.0 \times 10^{-2}$ is chosen
and the gap exists in $0.1475\rlc < s < 0.3818\rlc$.

It follows from the figure that the solid line 
matches the observed flux with a reasonable transfield thickness,
$D_\perp=0.61 s^{\rm in} = 0.38 W$.
On the other hand, for the dashed line, 
we have to choose $D_\perp=0.25\rlc = 1.7 s^{\rm in}$,
which is unrealistic.
The observed fluxes cannot be explained with acceptable
gap width (e.g., $D_\perp < s^{\rm in}$) 
no matter what we may adjust
$j^{\rm in}$, $j^{\rm out}$, and $j_{\rm gap}$
if $d=1.53$~kpc.

On these grounds, 
we conjecture that the distance $1.5\pm 0.4$~kpc
determined from the dispersion measure (Taylor \& Cordes 1993)
is too large 
and that a more closer distance, such as
$500$~pc derived from ROSAT data analysis 
($\ddot{\rm O}$gelman \& Finley 1993)
or $700$~pc estimated from a study of the extended nonthermal
radio source around the pulsar 
(Combi, Romero, Azc$\acute{\rm a}$rate 1997), 
is plausible.
This conclusion is consistent with the results obtained in Paper~IX
under the mono-energetic approximation 
that all the particles are saturated at
the equilibrium Lorentz factor,
which is justified for middle-aged pulsars.

\section{Discussion}
\label{sec:discussion}
In summary, we have quantitatively examined the stationary 
pair-production cascade in an outer magnetosphere,
by solving the set of Maxwell and Boltzmann equations 
one-dimensionally along the magnetic field.
We revealed that an accelerator (or a potential gap) 
is quenched by the created pairs in the gap
but is {\it not} quenched by the injected particles from outside
of the gap,
and that the gap position shifts as a function of the
injected particle fluxes:
If the injection rate across the inner (or outer) 
boundary approaches the typical Goldreich-Julian value,
the gap is located near to the light cylinder (or the star surface).
It should be emphasized that the particle energy distribution
is not represented by a power law,
as assumed in some of previous outer-gap models.
The particles escape from the gap with sufficient Lorentz factors
and emit significant photons in 100~MeV--3~GeV energies
via curvature radiation outside of the gap. 
The $\gamma$-ray spectrum including this component
explains the phase-averaged EGRET spectra for 
the Vela pulsar, the Geminga pulsar and PSR~B1055-52 
between 100~MeV and 6~GeV.
TeV fluxes are unobservable with current ground-based telescopes
for these pulsars.

We show that synchro-curvature process can be 
approximated by the pure-curvature one in the next subsection.
We then point out an implication to the $\gamma$-ray luminosity
versus the spin-down luminosity in \S~\ref{sec:lumin},
and discuss future extensions of the present method
in \S\S~\ref{sec:return}--\ref{sec:unif_pol}.

\subsection{Synchrotron vs. Curvature Processes}
\label{sec:small_pitch}
Let us first examine the pitch angles of particles injected across 
the inner boundary.
Imposing that the synchrotron cooling length scale,
\begin{equation}
  l_{\rm sync} 
  \equiv \frac{\Gamma m_{\rm e} c^2}{P_{\rm sync}/c}
  = \frac{3 m_{\rm e}^3 c^6}{2 e^4 B^2 \Gamma \sin^2\chi}
  \label{eq:cool_sync}
\end{equation}
is greater than the distance, $\varepsilon \rlc$,
for the particles to migrate before arriving the outer gap,
we obtain
\begin{equation}
  \sin\chi < 7.2 \times 10^{-6} 
             \left( \frac{\Omega_2}{\varepsilon_{0.1}\Gamma_2}
             \right)^{1/2}
             B_{10}^{-1},
  \label{eq:cool_pitch}
\end{equation}
where $\varepsilon_{0.1}\equiv \varepsilon/0.1$,
$\Gamma_2 \equiv \Gamma/10^2$,
$\Omega_2 \equiv \Omega/(10^2 \mbox{rad s}^{-1})$, and 
$B_{10} \equiv B/(10^{10} \mbox{G})$.
Particles having initial pitch angles greater than this value
will lose transverse momentum to reduce the
pitch angles below this value,
while migrating the distance $\varepsilon \rlc$.
Therefore, if the particles are supplied from the polar cap
with $\Gamma \sim 10^2$ for instance, we can safely state that the 
the pitch angles are less than $10^{-5}$ rad.

If a particle having such a small pitch angle is accelerated
to $\Gamma=10^7$ in a weak magnetic field region $B=10^6$~G, 
the ratio
between the synchrotron and the curvature radiation rate becomes
\begin{eqnarray}
 \lefteqn{ 
  \frac{P_{\rm sync}}{\Pcv}
  = \left( \frac{\rho_{\rm c} \sin\chi}
                  {\Gamma m_{\rm e}c^2/eB} \right)^2}
  \nonumber\\
  &=& 7.7 \times 10^{-3}
      \Omega_2^{-2} B_6^2 \Gamma_7^{-2} 
      \left(\frac{\rho_{\rm c}}{0.5\rlc} \cdot
            \frac{\sin\chi}{10^{-5}}
      \right)^2,
  \label{eq:PsyPcv}
\end{eqnarray}
where $B_6 \equiv B/10^6 \mbox{G}$,
$\Gamma_7 \equiv \Gamma/10^7$.
It follows that the synchro-curvature radiation
can be approximated by a pure-curvature one 
for the particles injected across the inner boundary.
It is noteworthy that the density of created particles
is much less than that of injected ones 
(i.e., $j_{\rm gap} \ll j^{\rm in}$);
therefore, the synchro-curvature effects for the created particles,
which have much greater pitch angles compared with the injected ones,
are negligibly small for the gap electrodynamics
as well for the resultant $\gamma$-ray spectrum.

\subsection{Gamma-ray vs. Spin-down Luminosities}
\label{sec:lumin}
It should be noted that 
the emission from the escaping particles attain typically 
$40\%$ of the total $\gamma$-ray luminosity for young pulsars.
Thus, it is worth mentioning its relationship 
with the spin-down luminosity,
\begin{equation}
  L_{\rm spin}= -I \Omega \dot{\Omega}  \propto \Omega^{n+1},
  \label{eq:spindown}
\end{equation}
where the braking index $n$ is related to the spin-down rate as
\begin{equation}
  \dot{\Omega}= -k \Omega^n.
  \label{eq:braking}
\end{equation}
If the spin down is due to the magnetic dipole radiation, 
we obtain $n=3$.

The outwardly propagating particles escape from the gap 
with spatial number density (eq.~[\ref{eq:def-n}])
\begin{equation}
  N_{\rm out}
  = (j^{\rm in}+j_{\rm gap})\frac{\Omega B^{\rm out}}{2\pi ce},
  \label{eq:NeOUT}
\end{equation}
where $B^{\rm out}=B(\xi^{\rm out})$.
Therefore, the energy carried by the escaping particles 
per unit time is given by
\begin{equation}
  L_{\rm esc}
  = D_\perp^2 c N_{\rm out} \Gamma_{\rm esc} m_{\rm e}c^2,
  \label{eq:def_Lesc}
\end{equation}
where $\Gamma_{\rm esc} (\sim 10^{7.5})$ refers to the Lorentz factor 
of escaping particles.
Note that $\Gamma_{\rm esc}$ is essentially determined by the
equilibrium Lorentz factor (dotted line in fig.~\ref{fig:EVela_75a})
near the gap center.
Since the equilibrium Lorentz factor depends on the one-fourth power
of $\Ell$, the variation of $\Gamma_{\rm esc}$ on pulsar parameters
is small.
We can approximate $B^{\rm out}$ as
\begin{equation}
  B^{\rm out} \sim \frac{\mu}{\rlc^3}  
     \left( \frac{\rlc}{r^{\rm out}} \right)^3,
  \label{eq:Bout}
\end{equation}
where $r^{\rm out}$ refers to the distance of the outer boundary 
of the gap from the star center.
Let us assume that the position of the gap with respect to 
the light cylinder radius,
$r^{\rm out}/\rlc$, 
does not change as the pulsar evolves; this situation can be realized if 
$j^{\rm in}-j^{\rm out}$ is unchanged.
Evaluating $B$ at $r=0.5\rlc$, we obtain
\begin{eqnarray}
  L_{\rm esc} 
  &=& \frac{4 \Gamma_{\rm esc} m_{\rm e}c}{\pi e} 
      \mu \Omega^2 \left(\frac{D_\perp}{\rlc}\right)^2
  \nonumber\\
  &\propto& L_{\rm spin}{}^{0.5},
  \label{eq:Lesc}
\end{eqnarray}
where $n=3$ is assumed in the second line.
To derive this conclusion,
it is essential that the particles are not saturated
at the equilibrium Lorentz factor.
Thus, the same discussion can be applied irrespective of the
gap position or the detailed physical processes involved.
For example, an analogous conclusion was derived for a polar-cap model
by Harding, Muslimov, and Zhang (2002).
It is, therefore, concluded that the observed relationship
$L_\gamma \propto L_{\rm spin}{}^{0.5}$ 
merely reflects the fact that the particles are
unsaturated in the gap
and does not discriminate the gap position.

Let us compare this result with what would be expected in 
the CHR picture.
Since the gap is extended significantly along the field lines 
in the CHR picture,
particles are saturated at the equilibrium Lorentz factor
to lose most of their energies within the gap,
rather than after escaping from it.
We can therefore estimate the $\gamma$-ray luminosity as
\begin{equation}
  L_{\rm gap}
  = (D_\perp D_\phi W)
    \cdot N_{\rm out}
    \cdot \Pcv
  \label{eq:def_Lgap}
\end{equation}
where $D_\phi$ refers to the azimuthal thickness of the gap,
$\Pcv$ [ergs s$^{-1}$ (particle)$^{-1}$]
represents the curvature radiation rate. 
Noting that the particle motion saturates at the equilibrium 
Lorentz factor satisfying $\Pcv/c=e(-d\Psi/ds)$,
recalling that the acceleration field is given by
$-d\Psi/ds \approx \Omega B D_\perp^2 / 4\rho_{\rm c}c$
in the CHR picture,
and evaluating $B$ at $r=\rlc$,
we obtain
\begin{eqnarray}
  L_{\rm gap}
  &=& \frac{\mu^2 \Omega^4}{4\pi c^3}
      \frac{D_\perp^3 D_\phi W}{\rlc^5}
      \left(\frac{\rho_{\rm c}}{0.5\rlc}\right)^{-1}
  \nonumber\\
  &\propto& L_{\rm spin}
  \label{eq:Lgap},
\end{eqnarray}
where $n=3$ is assumed again in the second line.
Even though the escaping particles little contribute 
to the $\gamma$-ray luminosity in the CHR picture, 
it is worth mentioning the work done by 
Crusius--W$\ddot{\rm a}$tzel and Lesch (2002),
who accurately pointed out 
the importance of the escaping particles
in the CHR picture,
when we interpret $L_\gamma \propto L_{\rm spin}^{0.5}$ relation.

As we have seen,
the particles being no longer accelerated
contribute for the $\gamma$-ray luminosity that is proportional 
to $L_{\rm spin}^{0.5}$.
Reminding that the particles migrate with larger Lorentz factors
than the equilibrium value in the outer part of the gap 
(see fig.~\ref{fig:EVela_75a}),
we can expect roughly half of the $\gamma$-ray luminosity
is proportional to $L_{\rm spin}^{0.5}$
(mainly between 100~MeV and 1~GeV),
and the rest of the half to $L_{\rm spin}$
(mainly above 1~GeV).
As a pulsar ages,
its declined surface emission results in a large 
pair-production mean free path, and hence $W$.
Because $\vert \rhoGJ \vert \propto r^{-3}$ becomes small 
in the outer part of such an extended gap,
$\Ell(\xi)$ deviates from quadratic distribution 
to decline gradually in the outer part (fig.~\ref{fig:EGemi_60}). 
As a result, particles tend to be saturated at the equilibrium value.
On these grounds, we can predict that
the $\gamma$-ray luminosity tends to be proportional to
$L_{\rm spin}$ with age, 
deviating from $L_{\rm spin}^{0.5}$ dependence for young pulsars.
 
In the present paper, we have examined the set of Maxwell and
Boltzmann equations
one-dimensionally both in the configuration and the momentum spaces
(i.e., only $\xi$ and $\Gamma$ dependences are considered.) 
In the next three sections,
we discuss the extension of the present method 
into higher dimensions

\subsection{Returning Particles}
\label{sec:return}
If we consider the pitch-angle dependence of particle
distribution functions,
we can compute the radiation spectrum
with synchro-curvature formula (Cheng and Zhang 1996).
Moreover, we can also consider the returning motion of particles
inside and outside of the gap.
The returning motion becomes particularly important
when both signs of charge are injected across the boundary.
For example, not only positrons but also electrons could be
injected across the inner boundary from the polar-cap accelerator.
If $\Ell>0$ for instance, the injected electrons return
in the gap.
This returning motion significantly affects the Poisson equation,
if their injection rate is a good fraction of 
the Goldreich-Julian value.

It remains an unsettled issue whether an outer-gap accelerator
resides on the field lines on which a polar-cap accelerator
exists.
To begin with, 
let us consider the case when the plasma flowing 
between the polar cap and the outer-gap accelerator 
is completely charge separated.
Such a situation can be realized, for instance, 
if only positively charged particles are ejected outwardly 
from the polar cap while
there is virtually no electrons ejected inwardly from 
the outer gap.
Neglecting the pair production, 
current conservation law gives the charge density, $\rho_{\rm e}$,
per unit magnetic flux tube as 
\begin{equation}
  \frac{\rho_{\rm e}}{B} \propto \frac{j_{\rm tot}}{v},
  \label{eq:sep_flow}
\end{equation}
where $v$ refers to the particle velocity along the field line,
and $j_{\rm tot}$ the conserved current density per magnetic flux tube.
At each point along the field line,
$\rho_{\rm e}$ should match $\rho_{\rm GJ}$.
If the field line intersects the null surface, 
$\rho_{\rm e}$ must vanish there;
this obviously violates the causality in special relativity.
Therefore, a stationary ejection of a completely charge-separated plasma
from the polar cap
can be realized only along the field lines between 
the magnetic axis and those intersecting the null surface
at the light cylinder.
On these grounds, it was argued that an outer-gap accelerator,
which is formed close to the last-open field line,
may not resides on the same field lines on which 
a polar-cap accelerator resides.
This has been, in fact, the basic idea that an outer gap will
not be quenched, 
because the particles ejected from the polar cap will flow
along the different field lines.
This idea was welcomed in outer-gap models,
because a gap has been considered to be 
quenched if the external particle injection rate
becomes comparable to the Goldreich-Julian value,
which was proved to be incorrect in this paper.

In general, however, the plasmas are not completely charge separated
and consist of both signs of charge (e.g., positrons and electrons).
Such a situation can be realized, for instance, 
if both charges are ejected outwardly from a polar-cap accelerator,
or if positively charged particles are ejected outwardly
from the polar cap while
electrons are ejected inwardly from the outer gap,
or if there is a pair production between the two accelerators.
In these cases, the velocities of both charges
will be adjusted so that both the current conservation and 
$\rho_{\rm e}=\rho_{\rm GJ}$ are satisfied at each point along the
field lines.
Therefore, it seems likely that a polar-cap accelerator and 
an outer-gap accelerator reside on the same field lines.

To examine if there is a stationary plasma flow between the polar cap
and the outer gap,
we must extend the present analysis
into two dimensional momentum space in the sense that
the pitch-angle dependence of 
the particle distribution functions is taken into account
in addition to the Lorentz factor dependence. 
For example, if both charges are ejected from the polar-cap accelerator,
electrons will return in the outer gap,
screening the original acceleration field in the gap,
and violating the original balance of $\rho_{\rm e}=\rhoGJ$
outside of the gap.
Because the returning motion of particles can be treated correctly
if we consider the pitch-angle evolution of the distribution functions,
and because the pair production is already taken into account,
our present method is ideally suited to investigate the plasma flows
and $\Ell$ distribution self-consistently inside and outside of the gap.

\subsection{Unification of Outer-gap Models}
\label{sec:unif_out}
In addition to the extension into a higher dimensional momentum space,
it is also important to extend the present method 
into a two- or three-dimensional configuration space.
In particular, determination of the perpendicular thickness,
$D_\perp$, 
is important to constrain gap activities.
There have been, in fact, some attempts to constrain $D_\perp$
in the CHR picture.
Since $-d\Psi/ds$ is proportional to $B D_\perp^2$, 
particles energies, and hence the $\gamma$-ray energies
increase with increasing $D_\perp$ (for a fixed $B$).
Zhang and Cheng (1997) constrained $D_\perp$, by
considering the condition that the $\gamma$-rays
cause photon-photon pair production in the gap.
Subsequently,
Cheng, Ruderman, and Zhang (2000) extended this idea into
three-dimensional magnetosphere and discussed phase-resolved 
$\gamma$-ray spectra for the Crab pulsar.
In addition, 
Romani (1996) discussed the evolution of the $\gamma$-ray emission
efficiency and computed the phase-resolved spectra for the Vela pulsar,
by assuming that $BD_\perp^2$ declines as $r^{-1}$.
However, in these works, screening effects due to 
pair production has not been considered;
thus, the obtained $D_\perp$, as well as the assumed gap position
along the magnetic field, are still uncertain.

On the other hand, in our approach (picture), 
$D_\perp$ is not solved but only adjusted 
so that the $\gamma$-ray flux may match the observations.
Therefore, the question we must consider next is 
to solve such geometrical and electrodynamical
discrepancies between these two pictures.
We can investigate this issue by extending the present method 
into higher spatial dimensions.

\subsection{Unification of Outer-gap and Polar-cap Models}
\label{sec:unif_pol}
Electrodynamically speaking, the essential difference 
between outer-gap and polar-cap accelerators 
is the value of the optical depth for pair production.
In an outer-gap accelerator, 
pair production takes place via $\gamma$-$\gamma$ collisions 
and its mean-free path is much greater than the light cylinder radius.
Therefore, a pair production cascade takes place gradually in the gap.
In such a gap, $\Ell$ is screened out 
by the \lq generalized Goldreich-Julian charge density',
$f_{\rm null}$ (eq.~[\ref{eq:def_fnull}]),
which increases outwards 
if $\mbox{\boldmath$\Omega$}\cdot\mbox{\boldmath$B$}>0$.
Since $f_{\rm null}$ is negative (or positive) in the
inner (or outer) part of the gap,
there is a surface on which the right-hand side of 
equation~(\ref{eq:Poisson_1Db}) vanishes, 
as long as the injected current density is less than the 
Goldreich-Julian value.
The gap is located around this \lq generalized null surface', 
which is explicitly defined in \S~2 of Paper~VII.

On the contrary, in a polar-cap accelerator, 
pair production takes place mainly via magnetic pair production, 
of which mean free path is much less than the star radius 
for a typical magnetic field strength ($B \sim 10^{12}$~G, say).
As a result, 
a pair production avalanche takes place in a limited region, 
which is called as the \lq pair formation front', in the gap 
(Fawley, Arons, \& Sharlemann 1977; 
 Harding \& Muslimov 1998, 2001, 2002;
 Shibata, Miyazaki, Takahara 1998, 2002;
 Harding, Muslimov, Zhang 2002).
In the pair formation front,
a small portion of the particles return to screen out $\Ell$. 
Such a returning motion 
can be self-consistently solved together 
with $\Ell$ by our present method, 
if we implement the magnetic pair production 
and the resonant IC scattering redistribution functions 
in the source terms of the particles' and $\gamma$-rays' 
Boltzmann equations.
We can execute the same advection-phase computation in CIP scheme; 
thus, all we have to do is to add these source terms 
in the non-advection-phase computation, 
which is not very difficult.
Since analogous boundary conditions 
(e.g., $\Ell=0$ for a space-charge limited flow) will be applied,
we expect the present method is also applicable 
to a polar-cap accelerator.
This is an issue to be examined in our subsequent papers.

\par
\vspace{1pc}\par

One of the authors (K. H.) wishes to express his gratitude to
Drs. K. Shibata and A. Figueroa-Vin{$\bar{\rm a}$}s 
for valuable advice on numerical analysis, 
and to Drs. K.~S.~Cheng and 
C. Thompson for fruitful discussion on theoretical aspects.
He also thanks Canadian Institute for Theoretical Astrophysics
for welcoming him as a visiting researcher.

\newpage
\appendix
\section{Reduction of Particle Boltzmann Equations}
\label{sec:red_boltz}
In this appendix, we derive the Boltzmann equations
~(\ref{eq:boltz_2}) and (\ref{eq:boltz_3})
from equation~(\ref{eq:boltz_1}).
To begin with, we consider the momentum derivative terms
$\mbox{\boldmath$F$}_{\rm ext} \cdot \partial_{\mbox{\boldmath$p$}} N$
in equation~(\ref{eq:boltz_1}).
Let us describe the momentum vector {\boldmath$p$} as
\begin{equation}
  \mbox{\boldmath$p$}
  = p_\parallel \mbox{\boldmath$e$}_\parallel
   +p_\phi      \mbox{\boldmath$e$}_\phi
   +p_z         \mbox{\boldmath$e$}_z,
  \label{eq:A06}
\end{equation}
where ($\mbox{\boldmath$e$}_\parallel$,
 $\mbox{\boldmath$e$}_\phi$,
 $\mbox{\boldmath$e$}_\perp$) 
forms the orthonormal basis:
$\mbox{\boldmath$e$}_\parallel$ and $\mbox{\boldmath$e$}_\perp$
are the unit vectors parallel and perpendicular to the local magnetic 
field on the poloidal plane,
and $\mbox{\boldmath$e$}_\phi$ is the azimuthal unit vector.
Introducing polar coordinates, 
we express the components as
\begin{eqnarray}
  p_\parallel
    &=& p \cos\chi,                 \nonumber \\
  p_\phi
    &=& p \sin\chi \cos\chi_\perp,  \nonumber \\
  p_z
    &=& p \sin\chi \sin\chi_\perp,
  \label{eq:A07}
\end{eqnarray}
to obtain
\begin{eqnarray}
  \frac{\partial}{\partial p_\parallel}
  &=& \cos\chi\frac{\partial}{\partial p}
     -\frac{\sin\chi}{p}\frac{\partial}{\partial \chi},
     \nonumber \\
  \frac{\partial}{\partial p_\phi}
  &=& \sin\chi\cos\chi_\perp \frac{\partial}{\partial p}
     +\frac{\cos\chi\cos\chi_\perp}{p}
                             \frac{\partial}{\partial \chi}
     \nonumber \\
  & & -\frac{\sin\chi_\perp}{p \sin\chi}
                             \frac{\partial}{\partial \chi_\perp},
     \nonumber \\
  \frac{\partial}{\partial p_z}
  &=& \sin\chi\sin\chi_\perp \frac{\partial}{\partial p}
     +\frac{\cos\chi\sin\chi_\perp}{p}
                             \frac{\partial}{\partial \chi}
     \nonumber \\
  & & +\frac{\cos\chi_\perp}{p \sin\chi}
                             \frac{\partial}{\partial \chi_\perp}.
  \label{eq:A08}
\end{eqnarray}
The external force acting on a particle can be expressed as
\begin{equation}
  \mbox{\boldmath$F$}_{\rm ext}
  = q \left[ \left(-\frac{d\Psi}{ds}\right) \mbox{\boldmath$e$}_\parallel
            +\frac{\mbox{\boldmath$v$}}{c} \times
                   \mbox{{\boldmath$B$}}_{\rm p}
      \right]
   - \frac{\Pcv}{c} \frac{\mbox{\boldmath$v$}}
                         {\vert\mbox{\boldmath$v$}\vert},
  \label{eq:A09}
\end{equation}
where $\mbox{\boldmath$v$} \equiv 
       \mbox{\boldmath$p$}/(\Gamma m_{\rm e})$,
and $q$ designates the charge on the particle.
Substituting equations~(\ref{eq:A07}) into equation~(\ref{eq:A09}),
and using equation~(\ref{eq:A08}), we obtain
\begin{eqnarray}
  \mbox{\boldmath$F$}_{\rm ext} 
  \cdot \frac{\partial N}{\partial \mbox{\boldmath$p$}}
  &=& \left[ q \cos\chi \left(-\frac{d\Psi}{ds}\right)
            -\frac{\Pcv}{c}
      \right] \frac{\partial N}{\partial p}
  \nonumber \\
  & & \hspace*{-1.0truecm}
     -q \left(-\frac{d\Psi}{ds}\right) 
        \frac{\sin\chi}{p}
        \frac{\partial N}{\partial\chi}
     -\frac{q B_{\rm p}}{p} \frac{\partial N}{\partial\chi_\perp}.
  \label{eq:A10}
\end{eqnarray}
The first term on the right-hand side
describes the energy dependence of $N$.


Next, we consider the first and the second terms 
in equation~(\ref{eq:boltz_1}).
Decoupling the toroidal velocity associated with the drift motion as
\begin{equation}
  \tilde{\mbox{\boldmath$v$}} 
  \equiv \mbox{\boldmath$v$}-\varpi\Omega_{\rm p}\mbox{\boldmath$e$}_\phi,
  \label{eq:def_Vnc}
\end{equation}
we obtain
\begin{eqnarray}
  \lefteqn{\frac{\partial{N}}{\partial t}
            +\frac{\mbox{\boldmath$p$}}{m_{\rm e}\Gamma}
             \cdot \mbox{\boldmath$\nabla$}N
           = \frac{\partial{N}}{\partial t}
            +\mbox{\boldmath$v$} \cdot \mbox{\boldmath$\nabla$}N}
  \nonumber\\
  & & \hspace*{-1.0 truecm} 
     =\frac{\partial N}{\partial t}
            +\Omega_{\rm p}\frac{\partial N}{\partial \phi}
     +\tilde{v}_\parallel \frac{\partial N}{\partial s}
     +\tilde{v}_z         \frac{\partial N}{\partial z}
     +\frac{\tilde{v}_\phi}{\varpi}
      \frac{\partial N}{\partial\phi}.
  \label{eq:A03}
\end{eqnarray}

It is worth noting 
that the $z$ and $\phi$ derivatives in equation~(\ref{eq:A03})
denote the advection of $N$ in the trans-field directions.
In the present paper, however,
we are interested in only $s$ dependence in the configuration space.
Thus, we further reduce the right-hand side of 
equation~(\ref{eq:A03}), by utilizing the fact that
particles are frozen-in.
Assuming a stationary magnetosphere as represented
in equation~(\ref{eq:stationary}),
and integrating equation~(\ref{eq:A03}) over the momentum space,
we obtain
\begin{equation}
  \int \left[ \frac{\partial{N}}{\partial t}
               +\mbox{\boldmath$v$} \cdot \nabla N \right]
           d^3\mbox{\boldmath$p$}
  = \int \left( \mbox{\boldmath$v$}_{\rm p} \cdot
                \mbox{\boldmath$\nabla$} N
               +\frac{\tilde{v}_\phi}{\varpi}
                \frac{\partial N}{\partial\phi}
         \right)
    d^3 \mbox{\boldmath$p$},
  \label{eq:A05b}
\end{equation}
where 
$ \tilde{\mbox{\boldmath$v$}}_{\rm p} \equiv 
  \tilde{v}_\parallel \mbox{\boldmath$e$}_\parallel
 +\tilde{v}_z         \mbox{\boldmath$e$}_z
$.

Let us introduce the averaged particle velocity such that
\begin{equation}
  \langle\tilde{\mbox{\boldmath$v$}}_{\rm p}\rangle
  \equiv \frac{\displaystyle \int \tilde{\mbox{\boldmath$v$}}_{\rm p} N 
               d^3\mbox{\boldmath$p$}}
              {\displaystyle \int N d^3\mbox{\boldmath$p$}},
  \quad
  \langle\tilde{v}_\phi\rangle
  \equiv \frac{\displaystyle \int \tilde{v}_\phi N 
               d^3\mbox{\boldmath$p$}}
              {\displaystyle \int N d^3\mbox{\boldmath$p$}}.
  \label{eq:A05c}
\end{equation}
Then noting (Bekenstein \& Oron 1978)
\begin{equation}
  \langle\tilde{\mbox{\boldmath$v$}}_{\rm p}\rangle
   =\pm \frac{\vert\langle\tilde{\mbox{\boldmath$v$}}_{\rm p}
                   \rangle\vert}{\Bp}
        \mbox{\boldmath$B$}_{\rm p},
  \quad
  \langle\tilde{v}_\phi\rangle
  =\pm \frac{\vert\langle\tilde{\mbox{\boldmath$v$}}_{\rm p}
                  \rangle\vert}{\Bp}
       B_\phi,
\end{equation}
where plus (or minus) sign is chosen for outwardly (or inwardly)
propagating particles,
we obtain
\begin{equation}
  \int \left[ \frac{\partial{N}}{\partial t}
               +\mbox{\boldmath$v$} \cdot \nabla N \right]
           d^3\mbox{\boldmath$p$}
  =\pm \vert\tilde{\mbox{\boldmath$v$}}_{\rm p}\vert
       \mbox{\boldmath$B$} \cdot \mbox{\boldmath$\nabla$}
       \left[ \frac{1}{\Bp} \int N d^3 \mbox{\boldmath$p$}
       \right],
  \label{eq:A05bb}
\end{equation}
where $\mbox{\boldmath$\nabla$} \cdot \mbox{\boldmath$B$}=0$ is used.
We can neglect the azimuthal derivative in equation~(\ref{eq:A05bb}),
if the azimuthal thickness is large compared with those on the poloidal
plane, 
or if $B_\phi$ is small compared with $\Bp$.
Under these assumptions, equation~(\ref{eq:A05bb}) reduces to
\begin{equation}
  \int \left[ \frac{\partial{N}}{\partial t}
               +\mbox{\boldmath$v$} \cdot \nabla N \right]
           d^3\mbox{\boldmath$p$}
  =\int \left[ \pm \vert\tilde{\mbox{\boldmath$v$}}_{\rm p}\vert
               \Bp \frac{\partial}{\partial s}
               \left(\frac{N}{\Bp}\right)
        \right]
   d^3 \mbox{\boldmath$p$}
  \label{eq:A05h}
\end{equation}

On these grounds, we neglect $z$ and $\phi$ dependence of the
distribution functions and approximate equation~(\ref{eq:A03}) as 
\begin{equation}
  \frac{\partial{N}}{\partial t}
           +\mbox{\boldmath$v$} \cdot \mbox{\boldmath$\nabla$}N
  = \vert\tilde{\mbox{\boldmath$v$}}_{\rm p}\vert
    \Bp \frac{\partial}{\partial s} \left(\frac{N}{\Bp}\right).
  \label{eq:A05i}
\end{equation}
Neglecting the $\chi_\perp$ derivative in equation~(\ref{eq:A10}),
and assuming 
$\vert\tilde{\mbox{\boldmath$v$}}_{\rm p}\vert=c \cos\chi$,
we finally obtain
\begin{eqnarray}
  & & \hspace*{-1.0truecm}
  c\cos\chi \Bp \frac{\partial}{\partial s}
               \left(\frac{N_\pm}{\Bp}\right)
    + \left[ \pm\cos\chi \cdot e \left(-\frac{d\Psi}{ds}\right)
            -\frac{\Pcv}{c}
      \right] \frac{\partial N_\pm}{\partial p}
  \nonumber \\
  &\mp& e \left(-\frac{d\Psi}{ds}\right) 
        \frac{\sin\chi}{p}
        \frac{\partial N_\pm}{\partial\chi}
       = S(s,p,\chi),
  \label{eq:A05j}
\end{eqnarray}
where $N_+$ (or $N_-$) denotes positronic (or electronic)
distribution function,
and $S$ refers to the source term averaged in a gyration.
If we neglect $\chi$ dependence,
we obtain equation~(\ref{eq:boltz_2}) and (\ref{eq:boltz_3}),
where $p=m_{\rm e}c\sqrt{\Gamma^2-1} \approx m_{\rm e}c\Gamma$
is used.
(It may be worth noting that
$\chi$ is different from the pitch angle
defined in the synchro-curvature formula derived by
Cheng and Zhang 1996, 
because the particles azimuthally drift,
while the synchro-curvature formula was derived 
when the guiding center moves {\it along} the magnetic field.)

\section{Inverse-Compton Scattering Redistribution Function}
\begin{figure*} 
\centerline{ \epsfxsize=12cm \epsfbox[0 0 450 250]{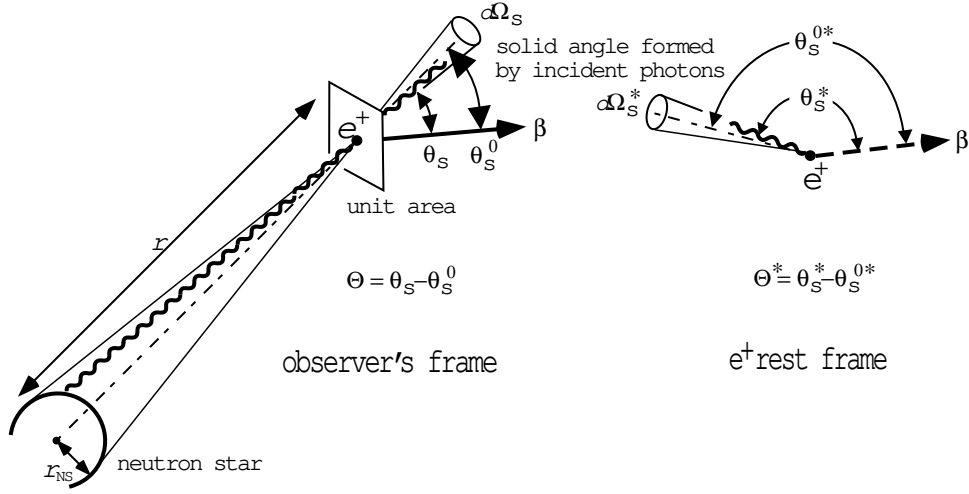} } 
\caption{
Scattering geometry in the observer's frame (non-primed)
and the positron rest frame (primed).
        }
\label{fig:ICframes}
\end{figure*}

In this appendix, we derive the redistribution function for
the IC scatterings.  
The number of photons upscattered by a single positron 
or electron into the energy interval $E_\gamma^\ast$ and
$E_\gamma^\ast+dE_\gamma^\ast$ in time interval $dt^\ast$
in the solid angle 
$d\Omega_\gamma^\ast(\theta_\gamma^\ast,\phi_\gamma^\ast)$
is given by
\begin{equation}
  dN_\gamma
  = dt^\ast \int dE_{\rm s}^\ast \frac{dF_{\rm s}^\ast}{dE_{\rm s}^\ast}
  \int\int \frac{d\sigma_{\rm KN}^\ast}
                {dE_\gamma^\ast d\Omega_\gamma^\ast}
           dE_\gamma^\ast d\Omega_\gamma^\ast,
\label{eq:B01}
\end{equation}
where the asterisk denotes that the quantity is measured in the 
positron (or electron) rest frame.
The incident photon flux per unit incident photon energy 
[photons s${}^{-1}$cm${}^{-2}$erg${}^{-1}$]
is given by
\begin{equation}
  \frac{dF_{\rm s}^\ast}{dE_{\rm s}^\ast}
  \equiv \frac{1}{E_{\rm s}^\ast}
         \int I_{\rm s}^\ast(E_{\rm s}^\ast,\Omega_{\rm s}^\ast) 
              \cos\Theta^\ast d\Omega_{\rm s}^\ast,
  \label{eq:B02}
\end{equation}
where $I_{\rm s}^\ast$ 
[ergs s${}^{-1}$cm${}^{-2}$ster${}^{-1}$erg${}^{-1}$]
is the specific intensity,
$\Theta^\ast$ the angle between the photon momentum and the normal
vector of a plane across which the incident flux is measured
(fig.~\ref{fig:ICframes}).

\begin{figure} 
\centerline{ \epsfxsize=12cm \epsfbox[-80 5 480 270]{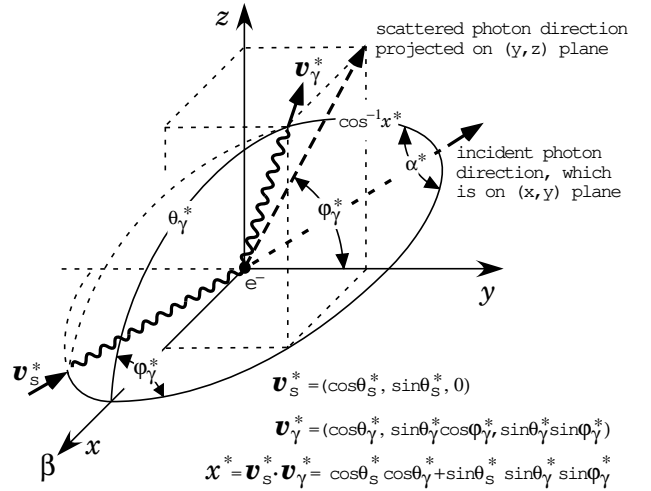} } 
\caption{
Definition of angles in the positron rest frame.
        }
\label{fig:ICrest}
\end{figure} 

When the particle is moving with Lorentz factor
$\Gamma=1/\sqrt{1-\beta^2}$, we can relate the particle rest frame
quantities (with asterisks) with the observer's frame 
(without asterisks) as
\begin{equation}
  E_{\rm s}^\ast = \Gamma(1-\beta\cos\theta_{\rm s})E_{\rm s},
  \quad 
  E_\gamma^\ast  = \Gamma(1-\beta\cos\theta_\gamma)E_\gamma,
  \label{eq:B03a}
\end{equation}
\begin{equation}
  d\Omega_{\rm s}^\ast
    = \frac{d\Omega_{\rm s}}{\Gamma^2(1-\beta\cos\theta_{\rm s})^2},
  \quad 
  d\Omega\gamma^\ast
    = \frac{d\Omega_\gamma}{\Gamma^2(1-\beta\cos\theta_\gamma)^2},
  \label{eq:B03b}
\end{equation}
\begin{equation}
  \cos\theta_{\rm s}^\ast
    = \frac{\cos\theta_{\rm s}-\beta}{1-\beta\cos\theta_{\rm s}},
  \quad 
  \sin\theta_{\rm s}^\ast
    = \frac{\sin\theta_{\rm s}}{\Gamma(1-\beta\cos\theta_{\rm s})}
  \label{eq:B03c}
\end{equation}
\begin{equation}
  \cos\Theta^\ast d\Omega_{\rm s}^\ast
  = \frac{\cos\Theta d\Omega_{\rm s}}
         {\Gamma^2(1-\beta\cos\theta_{\rm s})^2}.
  \label{eq:B03e}
\end{equation}
Moreover, we obtain the following Lorentz invariant,
\begin{equation}
  \frac{I_{\rm s}^\ast}{E_{\rm s}^\ast{}^3}
  = \frac{I_{\rm s}}{E_{\rm s}{}^3}.
  \label{eq:B04}
\end{equation}
In the observer's frame, the soft photon flux
[ergs s${}^{-1}$cm${}^{-2}$erg${}^{-1}$]
is given by
\begin{equation}
  E_{\rm s} \frac{dF_{\rm s}}{dE_{\rm s}}
  = \int I_{\rm s}(E_{\rm s},\Omega_{\rm s}) 
    \cos\Theta d\Omega_{\rm s}
  = \pi B_{\rm s}(T) \left(\frac{r_{\rm NS}}{r}\right)^2,
  \label{eq:B05}
\end{equation}
where $B_{\rm s}$ is the Planck function is defined by
equation~(\ref{eq:planck}).
Since 
\begin{equation}
  d\Omega^\ast \cos\Theta^\ast \frac{I_{\rm s}^\ast}{E_{\rm s}^\ast}
  = d\Omega \cos\Theta \frac{I_{\rm s}}{E_{\rm s}}
  \label{eq:B07}
\end{equation}
holds for $r \gg r_{\rm NS}$, we obtain
\begin{equation}
  \frac{dF_{\rm s}^\ast}{dE_{\rm s}^\ast}
  = \frac{dF_{\rm s}}{dE_{\rm s}}
  = \pi \frac{B_{\rm s}(T)}{E_{\rm s}}
        \left( \frac{r_{\rm NS}}{r} \right)^2,
  \label{eq:B08a}
\end{equation}
as long as the blackbody emission comes from the whole surface
of the neutron star.
If the observed emission area of the $i$th blackbody component
is $A_j 4\pi r_{\rm NS}^2$, equation~(\ref{eq:B08a}) is modified as
\begin{equation}
  \frac{dF_{\rm s}^\ast}{dE_{\rm s}^\ast}
  = \pi \left( \frac{r_{\rm NS}}{r} \right)^2
    \frac{1}{E_{\rm s}}
  \sum_j A_j B_{\rm s}(T_j)
  \label{eq:B08b}
\end{equation}

The IC redistribution function is defined by
\begin{equation}
  \etaICg(\Eg,\Gamma,\mu_{\rm s})
  \equiv \int d\Omega_\gamma 
              \frac{dN_\gamma}
                   {dt d(E_\gamma/m_{\rm e}c^2) d\Omega_\gamma}
  \quad \mbox{[s${}^{-1}$]},
  \label{eq:B09}
\end{equation}
where $\mu_{\rm s}\equiv \cos\theta_{\rm s}$.
Noting that 
\begin{equation}
  dt dE_\gamma d\Omega_\gamma
  = \Gamma^2(1-\beta\cos\theta_\gamma)
    dt^\ast dE_\gamma^\ast d\Omega_\gamma^\ast,
  \label{eq:B10}
\end{equation}
and assuming that the specific intensity of the incident photons are
unidirectional (i.e., $\mu_{\rm s}=\mu_{\rm s}^0=$constant), 
we obtain
\begin{eqnarray}
  \etaICg(\Eg,\Gamma,\mu_{\rm s}^0)
  &=& \frac{1-\beta\mu_{\rm s}^0}{\Gamma}
      \int \frac{d\Omega_\gamma^\ast}{1+\beta\cos\theta_\gamma^\ast}
  \nonumber \\
  &\times&  
  \int d\Es \frac{dF_{\rm s}}{d\Es}
            \frac{d\sigma_{\rm KN}^\ast}{d\Eg^\ast d\Omega_\gamma^\ast},
  \label{eq:B11}
\end{eqnarray}
where 
\begin{eqnarray}
  \frac{dF_{\rm s}}{d\Es}
  &\equiv& 
         \frac{2\pi(m_{\rm e}c^2)^3}{h^3 c^2}
         \left( \frac{r_{\rm NS}}{r} \right)^2
         \Es^2
         \sum_j \frac{A_j}{\exp(\Es/\delta_j)-1},
  \nonumber\\
  \delta_j &\equiv& \frac{kT_j}{m_{\rm e}c^2}; \quad
  \Es \equiv \frac{E_{\rm s}}{m_{\rm e}c^2}, \quad
  \Eg \equiv \frac{E_\gamma }{m_{\rm e}c^2}.
  \label{eq:B13}
\end{eqnarray}
In the particle rest frame,
the Klein-Nishina cross section is given by
\begin{eqnarray}
  \frac{d\sigma_{\rm KN}^\ast}{d\Eg^\ast d\Omega_\gamma^\ast}
  &\equiv& \frac{3\sgT}{16\pi}
           \left(  \frac{\Eg^\ast}{\Es^\ast} \right)^2
           \left(  \frac{\Es^\ast}{\Eg^\ast}
                  +\frac{\Eg^\ast}{\Es^\ast}
                  +(x^\ast)^2 -1 \right)
  \nonumber \\
  &\times&  \delta
            \left[ \Eg^\ast -\frac{\Es^\ast}{1+\Es^\ast(1-x^\ast)}
            \right],
  \label{eq:B14}
\end{eqnarray}
where 
\begin{equation}
  \Es^\ast \equiv \frac{E_{\rm s}^\ast}{m_{\rm e}c^2}, \quad
  \Eg^\ast \equiv \frac{E_\gamma^\ast }{m_{\rm e}c^2};
  \label{eq:B16}
\end{equation}
the quantity $x^\ast$ refers to the angle 
between the incident and the scattered
photon momenta (fig.~\ref{fig:ICrest}),
and can be expressed as
\begin{equation}
  x^\ast 
  = \cos\theta_{\rm s}^\ast \cos\theta_\gamma^\ast
   +\sin\theta_{\rm s}^\ast \sin\theta_\gamma^\ast \cos\phi_\gamma^\ast.
  \label{eq:def_xast}
\end{equation}

We define $S^\ast$ to be the plane containing both the incident and
scattered photon momenta.
Then $\alpha^\ast$ means the angle between the normal of $S^\ast$
and z axis.
Denoting the normal vector as
\begin{equation}
  \mbox{\boldmath$h$} = (h_x,h_y,\sqrt{1-h_x^2-h_y^2})
  \label{eq:def_h}
\end{equation}
and imposing
\begin{eqnarray}
  \mbox{\boldmath$h$} \cdot \mbox{\boldmath$v$}_{\rm s}^\ast
  &=& h_x \cos\theta_{\rm s}^\ast
     +h_y \sin\theta_{\rm s}^\ast
     = 0,
  \nonumber \\
  \mbox{\boldmath$h$} \cdot \mbox{\boldmath$v$}_\gamma^\ast
  &=& h_x \cos\theta_\gamma^\ast
     +h_y \sin\theta_\gamma^\ast \cos\phi_\gamma^\ast
  \nonumber \\
  &+& \sqrt{1-h_x^2-h_y^2} \sin\theta_\gamma^\ast \sin\phi_\gamma^\ast
     = 0,
  \label{eq:B16b}
\end{eqnarray}
we obtain
\begin{equation}
  \cos\alpha^\ast
  = \sqrt{1-h_x^2-h_y^2}
  = \frac{f_\Theta}{\sqrt{f_\Theta{}^2+\sin^2\phi_\gamma^\ast}},
  \label{eq:B16c}
\end{equation}
where $ f_\Theta
        \equiv \sin\theta_{\rm sp}^\ast \cot\theta_\gamma^\ast
              -\cos\theta_{\rm s}^\ast \cos\phi_\gamma^\ast$.
Solving equation~(\ref{eq:def_xast}) for $\cos\phi_\gamma^\ast$
and eliminating $\phi_\gamma^\ast$ in $f_\Theta$,
we obtain
\begin{equation}
  \cos\phi_\gamma^\ast
  = \frac{\theta_\gamma^\ast-x^\ast\cos\theta_{\rm s}^\ast}
         {\sin\theta_{\rm s}^\ast\sqrt{1-(x^\ast)^2}},
  \label{eq:B16e}
\end{equation}
which leads to
\begin{eqnarray}
  \cos\theta_\gamma^\ast
  &=& \sin\theta_{\rm s}^\ast \sqrt{1-(x^\ast)^2} \cos\alpha^\ast
    + x^\ast \cos\theta_{\rm s}
  \nonumber \\
  &\approx& 
  x^\ast\cos\theta_{\rm s}^\ast \quad \mbox{if $\Gamma \gg 1$}.
  \label{eq:B17}
\end{eqnarray}

Since $ d\Omega_\gamma^\ast
        = d(-\cos\theta_\gamma^\ast)d\phi_\gamma^\ast
        = dx^\ast d\alpha^\ast$ holds, 
we can change the integration variables to obtain
\begin{eqnarray}
  \lefteqn{ \etaICg(\Eg,\Gamma,\mu_{\rm s}^0)
            = \frac{1-\beta\mu_{\rm s}^0}{\Gamma}}
  \nonumber \\
  &\times&
  \int \frac{dx^\ast d\alpha^\ast}
            {1+\beta(\sin\theta_{\rm s}^0{}^\ast \sqrt{1-(x^\ast)^2} 
       \cos\alpha^\ast
       +x^\ast \cos\theta_{\rm s}^0{}^\ast)}
    \nonumber \\
  &\times&
  \int d\Es \frac{dF_{\rm s}(\Es)}{d\Es}
            \frac{d\sigma_{\rm KN}^\ast}
                 {d\Eg{}^\ast d\Omega_\gamma^\ast}.
  \label{eq:B19}
\end{eqnarray}
Here, $\Es^\ast$ and $\Eg^\ast$ in 
the right-hand side of equation~(\ref{eq:B14})
should be replaced with $\Es$ and $\Eg$ by equations~(\ref{eq:B03a}).

We define the dimensionless IC redistribution function by
equation~({\ref{eq:def_etaIC_0}).
Then, integrating over $\Eg$ between $b_{i-1}$ and $b_i$,
we obtain
\begin{eqnarray}
  \etaICg{}_{,i}(\Gamma)
  &=& \frac{3\sgT}{16\pi\omgp} (1-\beta\mu_{\rm s}^0)
      \int_{\Emin}^{\Emax} d\Es \frac{dF_{\rm s}}{d\Es}
      \int_{-1}^{1} dx^\ast
  \nonumber \\
  &\times&  
  \int_0^{2\pi} d\alpha^\ast 
     f_{\rm IC}(x^\ast,\alpha^\ast,\Es;\mu_{\rm s},\Gamma),
  \label{eq:B20}
\end{eqnarray}
where $f_{\rm IC}$ is defined by equation~(\ref{eq:def_fIC}).
Substituting $dF_{\rm s}/d\Es$,
and summing up the blackbody components, we obtain
equation~(\ref{eq:def_etaIC_2}),
where $\mu_{\rm s}^0{}^\ast$ and $\mu_{\rm s}^0$ are
denoted as $\mu_\ast$ and $\mu$ in equation~(\ref{eq:def_etaIC_2}).

\end{document}